\documentclass[numsec,webpdf,modern,small,namedate]{oup-authoring-template}%
\usepackage{xr-hyper}
\RequirePackage[OT1]{fontenc}
\RequirePackage{amsthm,amsmath,amsxtra}
\RequirePackage{natbib}
\RequirePackage{hypernat}
\RequirePackage{hyperref}
\hypersetup{colorlinks=true,citecolor=blue}

\usepackage{anyfontsize}
\renewcommand\normalsize{%
   \fontsize{6.3}{10}\selectfont 
}

\usepackage{colortbl,xcolor,makecell}
\usepackage{array}
\usepackage{mathrsfs}
\usepackage{amssymb,bbm}
\usepackage{amsfonts}
\usepackage{amsmath,amssymb}
\usepackage{booktabs}
\usepackage{graphicx}
\usepackage{graphics}
\usepackage{epsfig}
\usepackage{subfigure}
\usepackage{caption}
\usepackage{amscd}
\usepackage{color}
\usepackage{mathrsfs}
\usepackage{latexsym,bm}
\usepackage{array}
\usepackage{verbatim}
\usepackage{threeparttable}
\usepackage{algorithm}
\usepackage{algpseudocode}
\usepackage{multirow}
\usepackage{doi}

\numberwithin{equation}{section}
\theoremstyle{thmstyleone}%
\newtheorem{theorem}{Theorem}

\newtheorem{assumption}{Assumption}

\theoremstyle{thmstyletwo}%

\theoremstyle{thmstylethree}%

\makeatletter
\newcommand*{\addFileDependency}[1]{
  \typeout{(#1)}
  \@addtofilelist{#1}
  \IfFileExists{#1}{}{\typeout{No file #1.}}
}
\newcommand*{\rom}[1]{\expandafter\@slowromancap\romannumeral #1@}
\makeatother

\newcommand*{\myexternaldocument}[1]{%
    \externaldocument{#1}%
    \addFileDependency{#1.tex}%
    \addFileDependency{#1.aux}%
}
\myexternaldocument{Supp13_final}

\newcommand{\mbf}{\mathbf}
\newcommand{\mbb}{\mathbb}
\newcommand{\mrm}{\mathrm}
\newcommand{\mc}{\mathcal}
\newcommand{\mf}{\mathfrak}

\newcommand{\mi}{\mrm{i}}
\newcommand{\norm}{\mrm{Norm}}
\newcommand{\trun}{\mrm{Trun}}

\onecolumn

\graphicspath{{./fig/}}
\begin{document}

\journaltitle{Journal Title Here}
\DOI{DOI HERE}
\copyrightyear{2024}
\pubyear{2024}
\access{Advance Access Publication Date: Day Month Year}
\appnotes{Paper}

\firstpage{1}

\title[Spectral Change Point Detection]{Spectral Change Point Estimation for High Dimensional Time Series by Sparse Tensor Decomposition}

\author[1,$\ast$]{Xinyu Zhang\ORCID{0000-0002-0253-6138}}
\author[2]{Kung-Sik Chan}

\address[1]{\orgdiv{KLATASDS-MOE, School of Statistics, and Academy of Statistics and Interdisciplinary Sciences}, \orgname{East China Normal University}, \orgaddress{\street{Shanghai}, \postcode{200241}, \country{China}}}

\address[2]{\orgdiv{Department of Statistics and Actuarial Science}, \orgname{University of Iowa}, \orgaddress{\street{Iowa City}, \postcode{52246}, \state{IA}, \country{USA}}}

\corresp[$\ast$]{Xinyu Zhang, KLATASDS-MOE, School of Statistics, and Academy of Statistics and Interdisciplinary Sciences, East China Normal University, Shanghai, China. \href{Email:xyzhang@sfs.ecnu.edu.cn}{xyzhang@sfs.ecnu.edu.cn}}

\abstract{
Multivariate time series may be subject to partial structural changes over certain frequency band, for instance, in neuroscience. We study the change point detection problem with high-dimensional time series, within the framework of  frequency domain. The overarching goal is to locate  all change points and delineate which series are activated by the change, over which frequencies. 
In practice, the number of activated series per change and frequency could span from a few to full participation. We solve the problem by first computing a  CUSUM tensor based on spectra estimated from blocks of the time series.
A frequency-specific projection approach is applied for dimension reduction. 
The projection direction is estimated by a proposed tensor decomposition algorithm that  adjusts to the sparsity level of changes.
Finally, the projected CUSUM vectors across frequencies are aggregated for change point detection.
We provide theoretical guarantees on the number of estimated change points and the convergence rate of their locations.
We derive error bounds for  the estimated projection direction   for identifying the frequency-specific series activated in a change.
We provide data-driven rules for the choice of parameters.
The efficacy of the proposed method is illustrated by simulation, and applications in stock returns and seizure detection.
}


\maketitle

\section{Introduction}
In the context of high-dimensional time series, the stationary assumption is often suspect. 
Instead, data may be piecewise stationary, with the non-stationarity caused by structural breaks in the mean function
\citep{csorgo1997limit,fryzlewicz2014wild,jirak2015uniform,wang2018high, enikeeva2019highdimensional}, in the variance function \citep{aue2009break,kirch2015detection,wang2021optimal}, or in the  auto- and/or cross-covariance function (equivalently the spectral density function or simply referred to as the spectrum below). 
Here, we consider the problem of detecting and locating changes in the spectrum of high-dimensional time series. 

The spectrum provides comprehensive information about the dynamics  as it encapsulates both the contemporaneous and dynamic covariance structure of the time series components.
On the one hand, by incorporating the  autocovariances at all lags, it allows us to discern patterns and structural breaks that may not be evident when focusing solely on the contemporaneous covariances.
On the other hand, changes in a parametric model (e.g., VAR model) have been widely studied \citep{davis2006structural,chan2014group,kirch2015detection},
and such changes will also result in structure breaks in the spectrum.
Hence, frequency domain approaches are able to detect such changes, also 
offering the benefit of circumventing any model assumptions \citep{ombao2005slex,cho2012multiscale, adak1998timedependent,cho2024highdimensional}.
What's more, it is especially appealing  in fields where frequency  interpretations or periodic patterns are natural and crucial, e.g., electrical engineering, finance, physics, neuroscience, etc. 
Furthermore, from the frequency viewpoint, it is pertinent to identify over  which frequencies (CP frequencies) and across  which series (CP series) a structural change occurs.   
This problem of detecting frequency-series specific changes is underexplored, save two related works: 
\cite{schroder2019fresped} proposed the  FreSpeD method, which
detects change points based on one auto-spectrum (co-spectrum) at a time. \cite{preuss2015detection} considered the problem of change-point detection based on local  spectral density matrix, but  their method is not designed for high dimensional time series, nor frequency specific.

To simultaneously estimate the change point locations,  activated series and frequencies, we propose to explore the sparse spectral norm of the difference in the spectral matrix  per frequency for change point detection.
We focus on sparsity because in high dimension, changes tend to activate a few time series components, which poses a significant challenge, especially with weak signals.
In such cases,  aggregating information across all components via, say,  the spectral norm may lose power due to accumulation of error, whereas the proposed sparse method helps to capture and highlight the signals.
We emphasize that the proposed methodology is also adaptable to non-sparse settings, with the sparsity level determined in a data-driven way.
Similarly, changes may be sparse or dense over the frequencies.
For example, a covariance matrix shift generally affects all frequencies, while data from certain field, such as neuroscience, may exhibit changes only over specific frequency bands.
Thus, the proposed method employs frequency-specific detection followed by adaptive aggregation to address the multi-faceted nature of structural  changes in high-dimensional time series.

To achieve our goal, our methodology builds on  frequency-specific projection, with the projector obtained by a sparse tensor decomposition algorithm.
Specifically, we first sub-divide the data into blocks and compute their spectra \citep{adak1998timedependent}.   
We then transform the sequence of spectra per frequency into a third-order CUSUM tensor. 
How to integrate the information in the third-order tensor  for change point detection is a challenging problem.
Several  aggregation approaches for integrating information for change point detection have been proposed, including $l_2$-aggregation \citep{horvath2012changepoint}, $l_{\infty}$-aggregation \citep{jirak2015uniform}, among others. 
Considering our goals of specifying the CP series, we propose to use a sparse projection approach.
The projection approach has been used by \cite{wang2018high} and \cite{wang2021optimal} for change point detection respectively in the mean and variance function for independent data, where the projector is obtained through some matrix decomposition, in combination with the wild binary segmentation \citep{fryzlewicz2014wild} for multiple change points.
In comparison, our problem requires the development of novel algorithms, including a new variant of the tensor power method, and theory for extracting the signal from the third-order CUSUM tensor of time series data.
By exploring the special structures of the CUSUM tensor, we develop a novel special sparse tensor decomposition algorithm, based on the truncated matrix power and tensor power methods, to obtain the projection direction from the CUSUM tensor, and the algorithm is theoretically guaranteed.

With the projector, change point detection may be implemented based on the projected CUSUM in a frequency-specific manner.
However, an overall conclusion on the change point locations is often desired, the solution of which requires  integrating the information across frequencies. 
We do this with thresholded $l_1$ aggregation  \citep{cho2015multiplechangepoint} which achieves adaptivity across frequencies.
In sum, leveraging the tensor decomposition algorithm, we develop a novel change point detection approach with high dimensional time series, that can simultaneously identify which series and frequencies are activated per change.

We establish the convergence rate for the projector output by the proposed sparse tensor decomposition algorithm, via a perturbation analysis of the high dimensional spectral CUSUM tensor.
The significance of sparsity is evident as it results in a faster convergence rate. However, we emphasize that the proposed method can detect sparse to dense spectral changes,  via a data-driven sparsity parameter. 
Furthermore, we show that our method consistently estimates  both the number and locations of the change points, with the localization rate nearly achieving the minimax optimal rate for the covariance change point problem.

We  outline  the rest of the paper.
Section \ref{sec:main} elaborates the main problem.
We detail the proposed methods in Section \ref{sec:solution} with justification based on characterizing their theoretical properties via non-asymptotics in Section \ref{sec:theory}.
We discuss empirical issues for applying the proposed methods in Section \ref{subsec:para}, and report  the empirical performance of the proposed method via simulation in Section \ref{sec:simulation}. 
In Section \ref{sec:application}, we demonstrate the efficacy of the proposed methods by doing change point analyses with a panel of  S\&P100 returns and an EEG recording of neonatal seizure.
We conclude in Section \ref{sec:discussion}.
All proofs, more discussions and some additional simulation results are collected in the \emph{Supplementary Materials}.

\subsection{Notation}\label{subsec:notation}
We generally use bold-face upright letters (e.g., $\mbf{X}$) for matrices, bold-face italic letters (e.g., $\bm{\gamma}$) for vectors, 
 calligraphic-font letters (e.g., $\mc{F}$) for tensors, and superscript $^*$ to signify  tensor-induced vectors (e.g., $\mc{F}^*$).
For any real number $r$, $\lceil r \rceil$ denotes the ceiling function, i.e.,  the smallest integer  greater or equal to $r$,
and  $\lfloor r \rfloor=-\lceil -r \rceil$ denotes the floor function.
For any vector $\bm{c} \in \mbb{R}^p$,  $\|\bm{c}\|_0$ denotes its $l_0$ norm, $\|\bm{c}\|$  its Euclidean norm, and $\|\bm{c}\|_{\infty}$ its max norm.
For any $p \times p$  real matrix $\mbf{A}$,  
$\mbf{A}_{i,}$ ($\mbf{A}_{,i}$) denote its $i$-th row (column).
We follow the tensor convention in \cite{kolda2009tensor}.
Specifically, for a third-order tensor $\mc{G} \in \mbb{C}^{p_1 \times p_2 \times p_3}$ and some $1\leq j\leq p_2, 1\leq l \leq p_3$, the mode 1 fibre is given by $\mc{G}_{:jl}$, and the slice along mode 3 is given as $\mc{G}_{::l}$.
When $\mc{G}$ is real, define the vector product with $\bm{c}^{(k)} \in \mbb{R}^{p_{k}}$ with $k=1,2,3$, e.g.
$\mc{G} \times_{1} \bm{c}^{(1)} \equiv \sum_{i=1}^{p_1} \bm{c}_{i}^{(1)}\mc{G}_{i::}$.
For any set $\mbb{U}$, let $|\mbb{U}|$ be its cardinality.

\section{Main problem}\label{sec:main}

Let $\mbf{X}=(\mbf{X}_1,\ldots,\mbf{X}_N)\in \mbb{R}^{p\times N}$ be a $p$-dimensional centered multivariate time series of length $N$.
Let the scaled time be $t=n/N \in [0,1]$.
We assume that $\mbf{X}$ is a piecewise stationary process, i.e.,  
\begin{equation}\label{eq:linear}
    \mbf{X}_n=\sum_{m=0}^{\infty} \tilde{\mbf{A}}(v_q,m)\mbf{Y}_{n-m},\quad v_{q-1}<t=n/N\leq v_{q},
\end{equation}
where $v_0=0<v_1<\dots<v_Q<v_{Q+1}=1$ are change points of unknown numbers at unknown locations,  the $\mbf{Y}$'s are $p'$-dimensional doubly infinite i.i.d. innovations with zero mean and identity covariance matrix, and $\tilde{\mbf{A}}(v_q,m)$ is a sequence of $p \times p'$ real coefficient matrices.
By assuming structural changes in the filter $\tilde{\mbf{A}}(v_q,m)$, this facilitates a  general framework for structural breaks in any auto- and/or cross- second moments across all lags. 
This formulation depicts the dependence structure of $\mbf{X}$ and encompasses several common process, such as VARMA processes and factor processes.
Additionally, it aligns with the functional dependence framework \citep{wu2005nonlinear}, facilitating further theoretical analysis.
Note that the proposed framework presumes no changes in the mean function, or the data has been pre-processed to remove any such changes. The interesting problem of joint detection of changes in the first and second moments awaits future investigation.

Changes in  $\tilde{\mbf{A}}(v_q,m)$ are succinctly  encapsulated  by the time-varying spectral density matrix $\mbf{f}(t,\omega)$.
Specifically, denote the transfer function at frequency $\omega$ as $\mbf{A}(v_q,\omega)=\sum_{m=0}^{\infty}\tilde{\mbf{A}}(v_q,m) \exp(-\mrm{i}\omega m)$,
it follows that $\mbf{f}(t,\omega)$ is piecewise constant such that
\begin{equation*}
\mbf{f}(t,\omega)=\mbf{f}(v_q,\omega)\equiv\mbf{A}(v_q,\omega)\mbf{A}^H(v_q,\omega),\quad v_{q-1}<t\leq v_{q} \,\, \mbox{for} \,\, q=1,\ldots,Q+1, \,\, \omega \in (-\pi, \pi],
\end{equation*}
where the superscript $^H$ denotes the complex conjugate transpose.
Hence, the differences $\mbf{f}(v_{q+1},\omega)-\mbf{f}(v_q,\omega)$ quantify the  structural breaks. Our objective is to leverage this observation to devise a  change point detection algorithm. 
As previously discussed, in high dimensional time series, a  structural break may activate only  a subset of time-series components, the search of which may be effected by finding a sparse linear projection of the vector time series that maximizes the spectral changes; such a projection may be frequency-specific.  The basic algebraic operation is the following sparse spectral norm.
Let $1\leq s \leq p$ be a fixed scalar.  For any  Hermitian matrix $\mbf{B} \in \mbb{C}^{p \times p}$, its sparse  spectral norm $\rho(\mbf{B}, s)$ is defined as
\begin{flalign}\label{eq:rhodef}
	\rho(\mbf{B}, s)\equiv \sup_{\substack{ \bm{c} \in \mbb{R}^{p}, \|\bm{c}\|_0\leq s, \|\bm{c}\|=1  }}|\bm{c}^{\top}\mbf{B}\bm{c}|.
\end{flalign}
As $\mbf{B}$ is Hermitian,  $\rho(\mbf{B}, s)=	\rho(\mrm{Re}(\mbf{B}), s)$. 
Hence, applying the preceding sparse spectral norm to detecting  changes in  $\mbf{f}(t,\omega)$ is equivalent to detecting changes in its real part $\mrm{Re}[\mbf{f}(t,\omega)]$, which is known as the co-spectrum. The co-spectrum admits  the interesting interpretation that it is proportional to the  covariance matrix of the component of $\mbf{X}$ corresponding to the  frequency $\omega$, see Section 7.1 in \cite{brillinger2001time}.
The co-spectral increment between two consecutive time segments is given by 
\begin{equation}\label{eq:g}
\mbf{g}_q(\omega)\equiv \mrm{Re}[\mbf{f}(v_{q+1},\omega)-\mbf{f}(v_q,\omega)]\quad \mbox{for} \quad q=1,\ldots,Q, \,\, \omega \in (-\pi, \pi].
\end{equation}
These are real symmetric matrices, thereby admitting an  eigen-decomposition.
Specifically, let $\lambda_{qi}(\omega)$ be the real eigenvalues ordered in decreasing magnitude, and  $\bm{\gamma}_{qi}(\omega)$  the corresponding eigenvectors.
The validity of our proposed methods require the following assumptions.

\begin{assumption}\label{ass:x}
$\mbf{X}$ is generated by (\ref{eq:linear}). There exist positive constants $c>0$ and $h>2$ such that for all $m \geq 0$, 
\begin{align}
	\begin{split}
		&\max_{1\leq q \leq Q+1} \max_{1\leq i \leq p} \|\tilde{\mbf{A}}_{i,}(v_q,m)\|\leq c(1+m)^{-h},\\
		&\max_{1\leq q \leq Q+1} \left(\sum_{i=1}^{p'} \|\tilde{\mbf{A}}_{,i}(v_q,m)\|_{\infty}^2\right)^{1/2}\leq c(1+m)^{-h}.
	\end{split}
 \label{eqn:cond-max}
\end{align}
\end{assumption}

Assumption \ref{ass:x} requires the process $\mbf{X}$ to be weakly dependent up to an algebraic decay rate. 
This assumption ensures consistent spectrum estimation with high-dimensional time series,  under the framework of functional dependence \citep{zhang2021convergence}.

\begin{assumption}\label{ass:factor}
For each $q=1,\ldots,Q$,
there exists a non-empty set of frequencies,  $\mbb{F}_q$, such that for all $\omega \in \mbb{F}_q$,  the leading eigenvalue $|\lambda_{q1}(\omega)|$ of $\mbf{g}_q(\omega)$ and the gap $\Delta \lambda_q (\omega)=|\lambda_{q1}(\omega)|-|\lambda_{q2}(\omega)|$ are both non-degenerate with $|\lambda_{q1}(\omega)| > \underline{\lambda}>0$
and $\Delta \lambda_q (\omega)>\underline{\Delta \lambda}>0$.
The leading eigenvector  $\bm{\gamma}_{q1}(\omega)$ is sparse such that $\sup_{q=1,\ldots,Q,
\omega \in \mbb{F}_q}\|\bm{\gamma}_{q1}(\omega)\|_0\leq k_0 < p$.
Assume that there exist positive constants $\underline{f}$ and $\bar{f}$ such that for all $q$ and $\omega$,
\begin{equation}
	k_0\underline{f} \leq \rho(\mrm{Re}(\mbf{f}(v_q,\omega)),k_0)\leq k_0 \bar{f}.   \label{eqn:cond-max-r}
\end{equation}
\end{assumption}

Assumption \ref{ass:factor} imposes conditions on the sparse spectral norm of the spectral densities  $\mbf{f}(v_q,\omega)$ and their increments $\mbf{g}_q(\omega)$.
Note that the RHS of (\ref{eqn:cond-max-r}) ensues from (\ref{eqn:cond-max}), and the LHS typically holds since the spectral density matrix is generally positive definite.
These quantities are related in the following sense:
\begin{equation}\label{eq:klambda}
	 \underline{\Delta\lambda} \leq \underline{\lambda}\leq 2 \max_q \rho(\mrm{Re}(\mbf{f}(v_q,\omega)),k_0) \leq 2 k_0 \bar{f}.
\end{equation}
Assumption \ref{ass:factor} allows but does not require that
changes in $\mbf{f}(t,\omega)$ occur only over a subset of frequencies that are specific to the change point. 
That is, the spectrum may remain constant at some frequencies but change over time at others.
This relaxation could be useful in fields where specific frequency change is common, such as neuroscience.
The frequencies in $\mbb{F}_q$ are referred to as the change-point frequencies at the $q$-th break.
For each $\omega \in \mbb{F}_q$, we assume that only a set of series experience spectral change at $\omega$, and such series  are referred to as the CP series, the collection of which is denoted as $\mbb{S}_q^{\omega}$.
Furthermore, let $\mbb{F}=\{\mbb{F}_q, 1\leq q \leq Q\}$ and
$\mbb{S}=\{\mbb{S}_q^{\omega}, 1\leq q \leq Q, \omega \in \mbb{F}_q\}$  be the list of all CP frequencies and CP series, respectively.
We remark that while our methodology performs optimally in the sparse series case, which is demonstrated in Theorem \ref{thm:tensor}, it could be applied in non-sparse settings as well.
Our goal is to detect the number and location of the change points, as well as the corresponding CP frequencies and CP series ($\mbb{F}$ and $\mbb{S}$).

As a simple example to illustrate the problem and assumptions, and to motivate the methodology, let us consider a $p$ dimensional time series $\mbf{X}$ with a single change point at $v$, where a new factor is introduced.
Specifically,  $\mbf{X}_n=\mbf{Z}_{n}$ before $v$, where $\mbf{Z}_n$ follows a  VMA model; $\mbf{X}_n=\mbf{Z}_{n}+\bm{\gamma} W_n$ after $v$, where $W_n$ is an univariate factor MA process independent of $\mbf{Z}_{n}$ and $\bm{\gamma}$ is the $p$-dimensional factor loading with sparse elements.
This process clearly satisfies (\ref{eq:linear}).
Additionally, the co-spectral increment $\mbf{g}(\omega)$ equals to the spectrum of $\bm{\gamma} W_n$.
Specifically, $\mbf{g}(\omega)=f_{W}(\omega)\bm{\gamma} \bm{\gamma}^{\top}$ is a rank one matrix with the leading and only eigenvalue being $f_{W}(\omega)$, i.e., the spectral density of $W_n$ at $\omega$,  and the according eigenvector being the sparse $\bm{\gamma}$.
Hence, Assumption \ref{ass:factor} naturally holds.
To assess the matrix $\mbf{g}(\omega)$ by its quadratic form, it is clear that the optimal vector (projection) is $\bm{\gamma}$.
This motivates the methodology of finding change points in the time-varying co-spectrum sequences using a dimension reduction procedure  via projection obtained by some sparse decomposition algorithm.
This example has been used as one of the simulations in Section 
\ref{sec:simulation} to further demonstrate the idea and performance.
It should be noticed that our paper addresses more complex settings than this simple example, including multiple change points, frequency specific projection, dependent changes and so on.

\section{Methodology}\label{sec:solution}

Our change point detection method proceeds as follows.
First, in Section \ref{subsec:cusum}, the entire time series is partitioned into multiple blocks, over each of which a spectral density matrix estimate is computed.
A CUSUM-type test statistic for change-point detection is then constructed based on the sequence of spectral density matrix estimates, resulting in a tensor.
Then, in Section \ref{subsec:proj}, a frequency-specific projection approach is proposed to compress the CUSUM tensor into a vector at each frequency.
This projection approach is realized by a special tensor decomposition algorithm, which is introduced in Section \ref{subsec:tensor}.
In Section \ref{subsec:cp}, these frequency-specific pieces of information are  aggregated via a sparsified wild binary segmentation algorithm for implementing multiple change point detection.

\subsection{CUSUM of the block-based spectral matrices}\label{subsec:cusum}
The first step is to estimate the time-dependent spectral density matrix $\mbf{f}(t,\omega)$, which requires 
 some trade-off between spectral estimation accuracy and  temporal resolution of the change point location.
Similar to \cite{adak1998timedependent} and with no loss of generality, we partition the index set $\{1,\ldots,N\}$ into $B$ time blocks of equal length $L$, which are defined by the boundaries $\{n_b=Lb,b=0,...,B\}$ with $B=\lfloor N/L \rfloor$. 
The change point search is then  on the block time scale, with the change  block index set defined as $\mbb{U}=\{u_{1\leq q\leq Q}|u_q= \lceil v_q B \rceil \}$. Note that the search can be  fine-tuned  using sliding windows around any change point detected on the block scale, which we will discuss later.

Denote $\Omega$ as a set of frequencies over which we estimate the spectrum. For instance, it can be the set of  Fourier frequencies $2\pi l/L$ for $1\leq l \leq \lceil L/2 \rceil$, or a set of evenly spaced frequencies of fixed size, or  certain  frequency band of interest, as in neuroscience. 
Consider the $b$-th block indexed by time $n_{b-1}+1,\ldots,n_b$.
For any frequency $\omega \in \Omega$, define $\mc{F}(\omega)$ as a $p\times p\times B$ (population) spectrum tensor with $\mc{F}_{::b}(\omega)=\mbf{f}_b(\omega)$ defined as a weighted average of $\mbf{f}(v_q,\omega)$ with weights proportional to number of observations within $b$-th block falling into the $q$-th segment; see the formula in (\ref{eq:f}).
Then, $\hat{\mc{F}}(\omega)$ is defined as a $p\times p\times B$ smoothed periodogram tensor with 
\begin{equation}\label{eq:fhat}
  \hat{\mc{F}}_{::b}(\omega)=\hat{\mbf{f}}_{b}(\omega)=\frac{1}{2\pi}\sum_{m=-R}^R K\left(\frac{m}{R}\right) \hat{\mbf{\Sigma}}_{b}(m)\exp(-\mi \omega m),
\end{equation}
where $K(\cdot)$ is a kernel function (Bartlett  kernel in our implementation)
 with bandwidth $1\leq R<L$, and $\hat{\mbf{\Sigma}}_{b}(m)$ is the lag-$m$ autocovariance matrix for the $b$-th block, i.e.,
\[
\hat{\mbf{\Sigma}}_{b}(m)=\frac{1}{L}\sum_{n=n_{b-1}+1+m}^{n_b}\mbf{X}_{n-m}\mbf{X}_n^{\top} \text{ for } m \geq 0,
\text{ and  } \hat{\mbf{\Sigma}}_{b}(m)=\hat{\mbf{\Sigma}}_{b}^{\top}(-m) \text{ for } m < 0.
\]

For a generic $p\times p\times B$ spectrum or periodogram tensor $\mc{G}$, define the co-spectrum CUSUM transformation $\mc{C}_{s,e}:\mbb{C}^{p\times p\times B}\to \mbb{R}^{p\times p\times (e-s)}$ on the interval $\{(s,e):1\leq s\leq e \leq B\}$ such that the $(b-s+1)$-th slice of $\mc{C}_{s,e}(\mc{G})$ for $b=s,\ldots, e-1$ is
\begin{align}\label{eq:that}
\mc{C}_{s,b,e}(\mc{G}) =\sqrt{\frac{(b-s+1)(e-b)}{e-s+1}}
\mrm{Re} \left(\frac{1}{e-b}\sum_{i=b+1}^{e} \mc{G}_{::i}-
\frac{1}{b-s+1}\sum_{i=s}^b \mc{G}_{::i}\right).
\end{align}
Specifically, when $s=1$ and $e=B$, for $b=1,\ldots, B-1$ we have
\begin{equation}\label{eq:t1b}
\mc{C}_{1,b,B}(\mc{G}) =\sqrt{\frac{b(B-b)}{B}}
\mrm{Re} \left(\frac{1}{B-b}\sum_{i=b+1}^B \mc{G}_{::i}-
\frac{1}{b}\sum_{i=1}^b \mc{G}_{::i}\right).
\end{equation}
Apply the $\mc{C}_{s,e}$ transformation to $\hat{\mc{F}}(\omega)$ and $\mc{F}(\omega)$, and for simplicity denote $\hat{\mc{T}}_{s,e}(\omega)=\mc{C}_{s,e}(\hat{\mc{F}}(\omega))$ and $\mc{T}_{s,e}(\omega)=\mc{C}_{s,e}(\mc{F}(\omega))$, and their $b$-th slices  as
$\hat{\mc{T}}_{s,b,e}(\omega)$ and $\mc{T}_{s,b,e}(\omega)$.

In the Supplementary Section \ref{sec:spectheory}, we provide non-asymptotic bounds on the deviation between  $\hat{\mc{T}}_{s,e}(\omega)$ and $\mc{T}_{s,e}(\omega)$ uniformly over $s,e$ and $\omega$, which facilitates deriving the theoretical properties of the change point detection method.

\subsection{A frequency-specific projection approach}\label{subsec:proj}

To extract and aggregate the information about the change points as contained in the third-order tensor $\hat{\mc{T}}_{s,e}(\omega)$ for fixed $\omega$,
we develop a projection approach to compress it into a projection vector along which the projected univariate series preserves the greatest change point signal.  Consider the simple case that the spectral density function changes at a frequency set $\Omega$ across a single change point at $v$ (i.e. $Q=1$).
For a specific frequency $\omega \in \Omega$, denote $v(\omega)=v$, and $u(\omega)$  the corresponding  change point block index. 
Let $\mbf{g}(\omega)=\mrm{Re}[\mbf{f}(v_2,\omega)-\mbf{f}(v_1,\omega)]$ be the co-spectral increment.

\begin{figure}
\centering
\includegraphics[width=\textwidth]{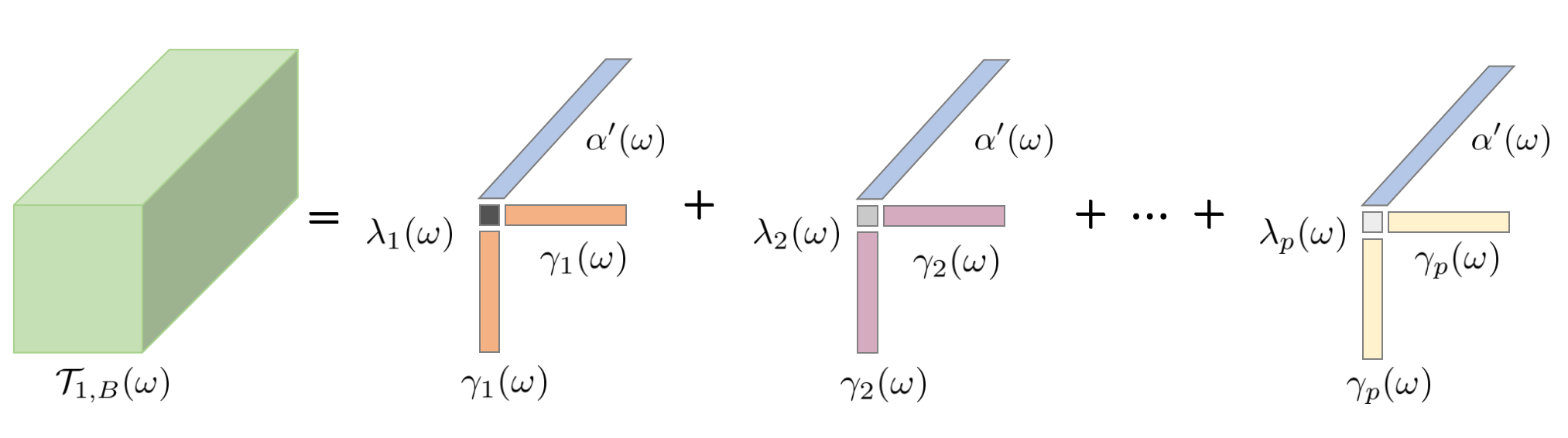}
\caption{A graphical illustration of (\ref{eq:t}).}
\label{fig:tensor}
\end{figure}

For notation simplicity, here assume that any change point is the first time point of some block (when change point is in the interior of some time blocks, just replace $u(\omega)$ below by $B v(\omega)$). Under this piecewise stationary structure, each slice of the co-spectrum CUSUM tensor in (\ref{eq:t1b}) can be computed explicitly as
\begin{equation}\label{eq:t1bb}
\mc{T}_{1,b,B}(\omega)=\left\{
\begin{array}{ll}
\sqrt{\frac{b}{B(B-b)}}(B-u(\omega))\mbf{g}(\omega),& \quad b< u(\omega),\\
\sqrt{\frac{(B-b)}{Bb}}u(\omega) \mbf{g}(\omega),& \quad b\geq u(\omega),\\
\end{array}
\right.
\end{equation}
which is product of the matrix $\mbf{g}(\omega)$ and a scalar quantity depending on location $b$.
Let 
\begin{flalign}\label{eq:alpha}
	\begin{split}
{\bm{\alpha}}'(\omega) \equiv \sqrt{\frac{1}{B}}\Big(&\sqrt{\frac{1}{B-1}}
(B-u(\omega)), \sqrt{\frac{2}{B-2}}(B-u(\omega)), \ldots,\\
&\sqrt{u(\omega)(B-u(\omega))}, 
\sqrt{\frac{B-u(\omega)-1}{u(\omega)+1}}u(\omega), \ldots, \sqrt{\frac{1}{B-1}}u(\omega)\Big)^{\top}
\end{split}
\end{flalign}
be the vector composed of all scalars
and let $\bm{\alpha}(\omega)=\norm({\bm{\alpha}}'(\omega))$, where $\norm(\bm{c})\equiv \bm{c}/\|\bm{c}\|$ is defined as the normalization operator on vector $\bm{c}$.
Let $\circ$ denotes the outer product.
Since (\ref{eq:t1bb}) could be represented as $\mc{T}_{1,b,B}(\omega)={\bm{\alpha}}'_b(\omega)\mbf{g}(\omega)$, combining this fact with the eigen-decomposition of $\mbf{g}(\omega)$ in Assumption \ref{ass:factor}, the tensor could be written as
\begin{equation}\label{eq:t}
\mc{T}_{1,B}(\omega)= \mbf{g}(\omega)\circ {\bm{\alpha}}'(\omega)
=\sum_{i=1}^p \lambda_i(\omega)\|{\bm{\alpha}}'(\omega)\|\cdot \bm{\gamma}_i(\omega)\circ\bm{\gamma}_i(\omega)\circ\bm{\alpha}(\omega).
\end{equation}
Thus, it is a special tensor in that it has a CANDECOMP/PARAFAC-decomposition whose modes 1 and 2 are 
symmetric and orthogonal, and identical mode 3 across the  components; see Figure \ref{fig:tensor} for 
a graphical illustration.

For the third-order tensor $\hat{\mc{F}}(\omega)$, we seek a projector projecting each of its frontal slice matrices into a scalar, thereby rendering  the tensor into a vector for change point detection.
The projected series should preserve the information about the change point as much  as possible.
Consider a real projector $\bm{\beta}(\omega) \in \mbb{R}^p$ and $\|\bm{\beta}(\omega)\|=1$ such that the tensor $\hat{\mc{F}}(\omega)$ is projected into a vector $ \hat{\mc{F}}(\omega) \times_1 \bm{\beta}(\omega) \times_2 \bm{\beta}(\omega)$, on which we could apply the operator $\mc{C}_{s,e}$  since vectors are $1\times1\times B$ tensors.
Meanwhile ${\mc{F}}(\omega)$ is projected into a vector $ {\mc{F}}(\omega) \times_1 \bm{\beta}(\omega) \times_2 \bm{\beta}(\omega)$, whose CUSUM will achieve the maximum magnitude with the projection  $\bm{\beta}(\omega)=\bm{\gamma}_1(\omega)$, and the maximum is 
\begin{align*}
\|\mc{T}_{1,B}(\omega)\times_1 \bm{\gamma}_1(\omega) \times_2 \bm{\gamma}_1(\omega)\|=\|\lambda_1(\omega){\bm{\alpha}}'(\omega)\|,
\end{align*}
hence retaining the most signal about the change point.
Thus, the oracle projection direction is the leading mode-1 component in the decomposition of the tensor $\mc{T}_{1,B}(\omega)$.
And to account for possible sparsity in the change series, the vector $\bm{\gamma}_1(\omega)$ is allowed to be sparse, as in Assumption \ref{ass:factor}.


\subsection{Special tensor decomposition}\label{subsec:tensor}

Below, we work on recovering the optimal projection $\bm{\gamma}_1(\omega)$ from the tensor  $\hat{\mc{T}}_{1,B}(\omega)$.
A state-of-art tensor decomposition technique is the alternating least squares (ALS) procedure \citep{kolda2009tensor}.
This involves solving the least squares problem one tensor mode at a time, while keeping the other modes fixed, and alternating between the tensor modes. 
Another approach is the tensor power method, which is basically a greedy or rank-1 ALS update.
This is a natural generalization of the matrix power method for extracting the dominant eigenvector, with a truncated version given by \cite{yuan2013truncated} for the sparse setting.
In sparse tensor decomposition, the tensor power method can be combined with a soft thresholding operator \citep{allen2012sparse} or a truncation step \citep{sun2017provable}, and the latter provided theoretical guarantees.
Since the tensor decomposition for ${\mc{T}}_{1,B}(\omega)$ enjoys the special features including partial orthogonality, symmetry and identical third mode, we utilize these features and develop a new Algorithm \ref{alg:tensor}.    
It combines the tensor power method with the truncated matrix power method.
For any  $\bm{c}\in \mbb{R}^p$, $\trun(\bm{c},k)$ equals the vector resulted from retaining its largest $k$ elements in magnitude but replacing other elements of $\bm{c}$ by zero. 
In Algorithm \ref{alg:tensor}, we first discard the boundary frontal slices of the tensor $\hat{\mc{T}}_{s,e}(\omega) \in \mbb{R}^{p\times p\times (e-s)}$,  for  technical reasons.
Specifically, let $\hat{\mc{T}}_{s,e}^{\dagger}(\omega) \in \mbb{R}^{p\times p\times (e-s-2\nu)}$ be its trimmed version by discarding $\nu$
slices at each boundary, based on which change point detection is proceeded.
The choice of $\nu$ is provided in Section \ref{sec:theory}.
For the single-change-point setting, Algorithm \ref{alg:tensor} is implemented with $s=1$ and $e=B$.

\begin{algorithm}[!htb]
\caption{Tensor power with truncated matrix power algorithm.}
\label{alg:tensor}
\linespread{1.35}\selectfont
\begin{algorithmic}[1]
\Require $\hat{\mc{T}}_{s,e}(\omega) \in \mbb{R}^{p\times p\times (e-s)}$, $k_0\leq k \leq p$
\Ensure $\hat{\bm{\gamma}}_{s,e}(\omega)=\bm{\gamma}^{(j)}(\omega)$
\State Let $j=1$ and initialize the unit vector $\bm{\gamma}^{(0)}(\omega) \in \mbb{R}^p$
\Repeat 
\State Compute $\bm{\alpha}^{(j)}(\omega)=\norm(\hat{\mc{T}}_{s,e}^{\dagger}(\omega) \times_{1} {\bm{\gamma}}^{(j-1)}(\omega) \times_{2} {\bm{\gamma}}^{(j-1)}(\omega))$
\State Compute $\mbf{D}^{(j)}(\omega)=\hat{\mc{T}}_{s,e}^{\dagger}(\omega) \times_{3} {\bm{\alpha}}^{(j)}(\omega)$
\State Let $i=1$ and $\bm{\gamma}^{(j,0)}(\omega)=\bm{\gamma}^{(j-1)}(\omega)$
\Repeat
\State Compute $\bm{\gamma}^{(j,i)}(\omega)=\norm(\trun(\norm(\mbf{D}^{(j)}(\omega)\bm{\gamma}^{(j,i-1)}(\omega)),k))$, update $i \leftarrow i+1$
\Until Convergence
\State Let $\bm{\gamma}^{(j)}(\omega)=\bm{\gamma}^{(j,i)}(\omega)$, update $j \leftarrow j+1$
\Until Convergence
\end{algorithmic}
\end{algorithm}

Algorithm \ref{alg:tensor} is a tensor power method, with $\bm{\alpha}^{(j)}(\omega)$ and $\bm{\gamma}^{(j)}(\omega)$ alternately updated according to lines 3 and 9.
Specifically, $\bm{\gamma}^{(j)}(\omega)$ is updated by a truncated matrix power method \citep{yuan2013truncated}, under the special case here that the matrix $\mbf{D}^{(j)}(\omega)$ is dynamic and updated along the outer iteration.
The iteration may be  terminated by keeping a pre-specified maximal number of iterations, or if the  changes in the $\bm{\alpha}^{(j)}(\omega)$'s and $\bm{\gamma}^{(j)}(\omega)$'s are below some threshold.
Let $K_{o}$ and $K_i$ be the number of outer and inner iterations in Algorithm \ref{alg:tensor}, respectively.
Then, the complexity of Algorithm 1 is $K_{o}\{k^2B+K_{i}(kp+p)\}$, with $O(k^2B)$ for the tensor power update (Lines 3 ad 4), and $O(kp+p)$ for the truncated matrix power update (Line 7), which consists of $O(kp)$ for matrix-vector product and $O(p)$ for selecting the top $k$ largest elements.
Although there are two levels of iterations, the outer iteration for $\bm{\alpha}^{(j)}(\omega)$ converges very quickly thanks to the identical third mode, thus a few outer iterations would work, which we explain in Supplementary Remark \ref{rem:alpha}.
Additionally, a good initialization $\bm{\gamma}^{(0)}(\omega)$ will further accelerate convergence, for which we provide several approaches
in Supplementary Section \ref{sec:initial}, using tensor slice aggregation or tensor unfolding.
The convergence of $\bm{\gamma}^{(j)}(\omega)$ output by this algorithm is theoretically guaranteed  by Theorem \ref{thm:tensor}, where the benefit of exploring the sparsity structure is also illustrated.

\subsection{Change points estimation}\label{subsec:cp}

Now, in the single-change-point case, the following projection procedure in Algorithm \ref{alg:single} could be applied with $s=1$ and $e=B$.
Specifically, the projector $\hat{\bm{\gamma}}_{s,e}(\omega)$ is estimated based on the CUSUM tensor $\hat{\mc{T}}_{s,e}(\omega)$, and further project it into a vector.
Note that the output $\hat{\sigma}_{s,e}(\omega)>0$ since $\hat{\mc{F}}(\omega)$ is positive semidefinite, and it accounts for the variation of the projected spectral density.
\begin{algorithm}[!htbp]
	\caption{Projection on CUSUM.}
	\label{alg:single}
	\linespread{1.35}\selectfont
	\begin{algorithmic}[1]
		\Require $\hat{\mc{F}}(\omega) \in \mbb{R}^{p\times p\times B}$, $k_0\leq k \leq p$, $s$, $e$
		\Ensure $\hat{\bm{\gamma}}_{s,e}(\omega)$, $\hat{\sigma}_{s,e}(\omega)$, $\hat{\mc{T}}^*_{s,e}(\omega)$
		\State Compute the CUSUM tensor  $\hat{\mc{T}}_{s,e}(\omega)$ by (\ref{eq:that}) with $\hat{\mc{F}}(\omega)$
		\State Implement Algorithm \ref{alg:tensor} with input $\hat{\mc{T}}_{s,e}(\omega)$ and $k$ to obtain $\hat{\bm{\gamma}}_{s,e}(\omega)$
		\State Obtain the projected  CUSUM vector $\hat{\mc{T}}^*_{s,e}(\omega)$ and a scaling factor $\hat{\sigma}_{s,e}(\omega)$ by
		\begin{align*}
		 \hat{\mc{T}}^*_{s,e}(\omega)
		&	\equiv\hat{\mc{T}}_{s,e}(\omega) \times_1 \hat{\bm{\gamma}}_{s,e}(\omega) \times_2 \hat{\bm{\gamma}}_{s,e}(\omega),\\
		\hat{\sigma}_{s,e}(\omega) &\equiv\frac{1}{e-s+1}\sum_{b=s}^e
		\hat{\mc{F}}_{::b}(\omega) \times_1 \hat{\bm{\gamma}}_{s,e}(\omega) \times_2 \hat{\bm{\gamma}}_{s,e}(\omega). 
		\end{align*}
	\end{algorithmic}
\end{algorithm}

In practice, the data may sustain multiple change points. 
Also, as previously discussed, an overall conclusion of change points across frequencies is  desired.
We combine the threshold $l_1$ aggregation \citep{cho2015multiplechangepoint} for multiple frequencies and wild binary segmentation  \citep{fryzlewicz2014wild} for multiple change points, which will be referred to as the  wild sparsified binary segmentation.
Intuitively, we apply wild binary segmentation to isolate individual change points within random windows over which we aggregate the information across frequencies. 
On a specific interval $(s,e)$, after obtaining $\hat{\mc{T}}^*_{s,e}(\omega)$ and $\hat{\sigma}_{s,e}(\omega)$ for each Fourier frequency $\omega \in \Omega$ by Algorithm \ref{alg:single}, 
let $\hat{\mc{F}}\in \mbb{R}^{p\times p\times B \times |\Omega|}$ be a 4-order tensor with $\hat{\mc{F}}_{:::l}=\hat{\mc{F}}(\omega_l)$ for $1\leq l \leq |\Omega|$.
We define a thresholded sum of the CUSUM across all frequencies, as follows:
\begin{equation}\label{eq:thre}
	\mf{C}_{s,e}(\hat{\mc{F}})=\sum_{\omega \in \Omega}\frac{|\hat{\mc{T}}^*_{s,e}(\omega)|}{\hat{\sigma}_{s,e}(\omega)}
	\mbb{I}\left(\frac{|\hat{\mc{T}}^*_{s,e}(\omega)|}{\hat{\sigma}_{s,e}(\omega)}>\tau\right),
\end{equation}
where $\mbb{I}$ is the indicator function, and $\tau$ is the threshold. 
The threshold step reduces the influence of irrelevant noisy parts, and can consistently detect change points.

\begin{figure}[!htb]
\centering
\includegraphics[width=0.8\textwidth]{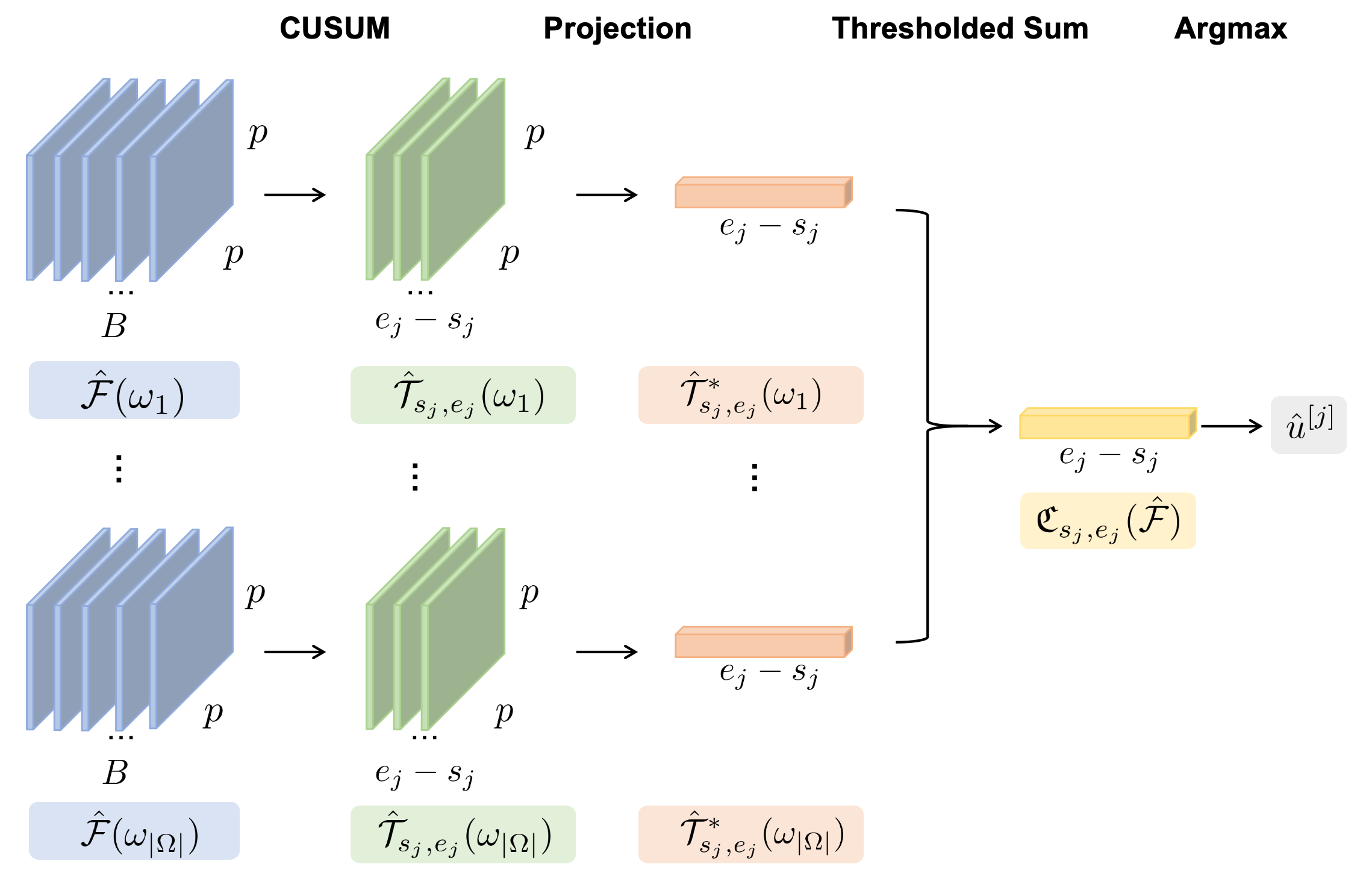}
\caption{A graphical illustration of Algorithm \ref{alg:multi} lines 6-9.}
\label{fig:alg}
\end{figure}

To implement the wild binary segmentation, we first randomly sample a large number of pairs, $(s_1,e_1),\ldots,(s_J,e_J)$, uniformly from the set $\{(s,e): 1\leq s <e \leq B\}$.
Then, for each $1\leq j \leq J$,  $\mf{C}_{s_j,e_j}(\hat{\mc{F}})$ is computed via  (\ref{eq:thre}).
Denote the maximizer of $\mf{C}_{s_j,e_j}(\hat{\mc{F}})$ over $[{s}_j+\nu, {e}_j-\nu]$ by  $\hat{u}^{[j]}$, and the corresponding maximum CUSUM value is $\mf{C}_{s_j,\hat{u}^{[j]},e_j}(\hat{\mc{F}})$.
Suppose $\mf{C}_{s_j,\hat{u}^{[j]},e_j}(\hat{\mc{F}}), 1\le j\le J$ is maximized at $j=\hat{j}$.
If $\mf{C}_{s_j,\hat{u}^{[\hat{j}]},e_j}(\hat{\mc{F}})$ is not above 0, conclude that  no change point is found, otherwise we declare  $\hat{u}^{[\hat{j}]}$  a change point and repeat the above procedure recursively on each segment to the left and right of $\hat{u}^{[\hat{j}]}$ to find other change points.
Thus, we carry out  change-point detection  recursively, with the search restricted to  those initial time windows $(s_j, e_j), 1\le j\le J$ that lie in an interval  under search.
The algorithm is detailed in Algorithm \ref{alg:multi}, with the key steps in lines 6-9 depicted in Figure \ref{fig:alg}.

\begin{algorithm}[!htb]
\caption{Multiple spectral change points detection algorithm.}
\label{alg:multi}
\linespread{1.35}\selectfont
\begin{algorithmic}[1]
\Require $\hat{\mc{F}} \in \mbb{R}^{p\times p\times B \times |\Omega|}$, $k_0\leq k \leq p$, $J \in \mbb{N}$, $\tau$
\Ensure $\hat{\mbb{U}}$
\State Set $\hat{\mbb{U}}=\emptyset$, randomly draw $(s_1,e_1),\ldots,(s_J,e_J)$, uniformly from $\{(s,e): 1\leq s <e \leq B\}$
\State Run \Call{SpecCp}{$\hat{\mc{F}},0,B$}
\Function{SpecCp}{$\hat{\mc{F}},s,e$}
\State Set $\mc{J}_{s,e} \equiv\{j: [s_j,e_j] \subseteq  [s,e], e_j-s_j+1> 2\nu\} \cup \{0\}$ where $s_0=s$ and $e_0=e$
\For {$j \in \mc{J}_{s,e}$ }
\For { $\omega \in \Omega$}
\State Run Algorithm \ref{alg:single} with $\hat{\mc{F}}(\omega)$, $k$, $s_j$, $e_j$ as input, and let $\hat{\bm{\gamma}}^{[j]}(\omega)$, $\hat{\sigma}_{s_j,e_j}(\omega)$, $\hat{\mc{T}}^*_{s_j,e_j}(\omega)$ be the output
\EndFor
\State Compute $\mf{C}_{s_j,e_j}(\hat{\mc{F}})$ by (\ref{eq:thre})
\EndFor
\State Let
$(j_0,\hat{u})=\arg\max_{j \in \mc{J}_{s,e}, {s}_j+\nu \leq u \leq {e}_j-\nu}\mf{C}_{s_j,u,e_j}(\hat{\mc{F}})$
\If {$\mf{C}_{s_{j_0},\hat{u},e_{j_0}}(\hat{\mc{F}})>0$}
\State Add $\hat{u}$ to the set $\hat{\mbb{U}}$
\State \Call{SpecCp}{$\hat{\mc{F}},s,\hat{u}$} and \Call{SpecCp}{$\hat{\mc{F}},\hat{u}+1,e$}
\EndIf
\EndFunction
\end{algorithmic}
\end{algorithm}

\section{Theoretical properties}\label{sec:theory}

In this section, we  establish the convergence of Algorithm \ref{alg:tensor} for sparse tensor decomposition and Algorithm \ref{alg:multi} for change point detections. 
We introduce some assumptions.

\begin{assumption}\label{ass:g}
	The $\mbf{Y}_{in}$'s are i.i.d. with zero mean and unit variance for $1\leq i \leq p$ and $1\leq n \leq N$. It follows either of the two conditions:
	(i) The $\mbf{Y}_n$'s are Gaussian random vectors with identity covariance matrix.
	(ii) For some $\iota>4$, $\mu_{\iota} \equiv \mbb{E}(|\mbf{Y}_{in}|^{\iota})<\infty$.
\end{assumption}

\begin{assumption}\label{ass:v}
	There exists $\delta>0$ such that 
	for all $q=1,\ldots,Q$, $\min(v_{q}-v_{q-1})\geq \delta$.
\end{assumption}

In Assumption \ref{ass:g}, the data are allowed to be Gaussianly distributed, or more generally, of finite polynomial moments, which permits heavy tail data.
Since the theory under (ii) involves more notations, we present it in the Supplementary Section \ref{sec:broadtheory}, and only show the theory under the Gaussian case (i) in the main file.
Simulations under Gaussian and non-Gaussian settings are also presented, respectively in the main file and Supplementary.
Under Assumption \ref{ass:g}(i), let the window size in (\ref{eq:fhat}) be $R \asymp(\frac{N \delta}{\log(Np)})^{1/3}$.
This choice is to the balance the bias variance trade-off in the CUSUM of the spectrum estimation.
Let the choice of boundary removal in Algorithm \ref{alg:tensor} and \ref{alg:multi} be $\nu\asymp{R \log(Np)}/{L}\leq \log(Np)$.
In Assumption \ref{ass:v},
if the number of change points is finite, then $\delta$  can be taken as some positive fraction,
otherwise, it decreases to zero with increasing $N$.

\begin{assumption}\label{ass:para}
	Let $\phi= \left(\frac{\log(Np)}{N}\right)^{1/3}$.  There exists a large enough absolute constant $c>1$  such that 
	(i) $\delta \geq c\phi^3$, 
	(ii) $\delta^{5/6} \underline{\Delta \lambda} \geq ck_0\sqrt{\phi}$,
	and 
	(iii) $c\left(\frac{N\delta}{\log(Np)}\right)^{1/3} \leq L \leq  N\delta$.
	
\end{assumption}

Assumption \ref{ass:para} (i) allows the high dimension setting with $p$ increasing with $N$ at some exponential rate so long as $\log(p)\leq \delta N-\log(N)$.
Also, as a result, the trimming satisfies $\nu \asymp{R \log(Np)}/{L} \lesssim \delta B$, hence  no change points near the boundary will be missed due to trimming. 
Assumption \ref{ass:para} (ii) imposes a lower bound on the signal strength by measuring the product $\delta^{5/6} \underline{\Delta \lambda}$. Note that if the time-series dimension $p$ increases sufficiently slowly as $N\to\infty$, for instance, if $\log(p)=o(N)$, then  $\phi\to 0$ and so does the lower bound.
Also, (iii) means that the block length $L$ should be smaller than the shortest inter-change-point distance but large enough to enable accurate per-block spectral density estimates. In sum, these are mild regularity assumptions.

Next we show the asymptotic validity of Algorithm \ref{alg:tensor} for the single-change-point setting, i.e. $s=1$ and $e=B$.

\begin{theorem}\label{thm:tensor}
	Suppose Assumptions \ref{ass:x}, \ref{ass:factor}, \ref{ass:g}(i), \ref{ass:v} and \ref{ass:para} hold.
	Let $Q=1$, $s=1$, $e=B$.
	Consider some frequency $\omega \in \mbb{F}_1$.
	Denote $\lambda(\omega)=|\lambda_1(\omega)|>0$.
	Assume $k$ satisfies
	\begin{equation}\label{eq:error}
		\left(1+2(k_0 / k)^{1 / 2}+2k_0 / k\right)\eta^2(\omega)<1,\quad \eta(\omega) \equiv \frac{\lambda(\omega)-(1-c_1)\Delta \lambda(\omega)}{\lambda(\omega)-c_1\Delta \lambda(\omega)}. 
	\end{equation}
	with some constant $0<c_1<1/2$.
	Assume the initial unit vector $\bm{\gamma}^{(0)}(\omega)$ is close enough to $\bm{\gamma}_1(\omega)$ such that $|\bm{\gamma}_1(\omega)^{\top}\bm{\gamma}^{(0)}(\omega)|$ is larger than some quantity defined in (\ref{eq:th1_gamma}).
	Let $a=2k+k_0$ with $k\gtrsim k_0$. 
	Then, with probability at least $1-N^{3-c_2}p^{2-c_2}$ for some $c_2>3$, we have
	\begin{equation}
		\sqrt{1-\left|\bm{\gamma}_1^{\top}(\omega)\hat{\bm{\gamma}}_{1,B}(\omega)\right|} \leq O\left(\frac{ a\phi }{\delta^{5/6}\underline{\Delta \lambda} }\right). \label{inner-prod-bound}
	\end{equation}
\end{theorem}

Under the condition that $a \asymp k_0$ is far less than $p$, the fact that this order grows linearly in $k_0$ illustrates the potential gain by taking into account of the sparsity in $\bm{\gamma}_1(\omega)$, which is corroborated by the simulation results reported below.
Assumption \ref{ass:para} (ii) implies that the RHS  (\ref{inner-prod-bound}) is further bounded by $\sqrt{\phi}$.
The consistency of the projector obtains if $\phi\to 0$, e.g., if $\log(p)=o(N)$; hence the projector casts light on the importance of each component series in the structure break.
As for the initialization conditions, Supplementary Section \ref{sec:initial} provides several implementation approaches and their theoretical discussions.

The following results state a non-asymptotic probability bound regarding the consistency of the proposed change-point detection method.
Let $\mbb{\hat{U}}$ denote the set of all change points  detected  by Algorithm \ref{alg:multi}, with   $\hat{Q}=|\mbb{\hat{U}}|$ equal to the number of  detected change points. Sort the  change point estimates (in ascending order) as $\hat{u}_q$ for $q=1,\ldots,\hat{Q}$.
Recalling that the trimming is $\nu\asymp{R \log(Np)}/{L}$.

\begin{theorem}\label{thm:cp}
	Suppose Assumptions \ref{ass:x}, \ref{ass:factor}, \ref{ass:g}(i), \ref{ass:v} and \ref{ass:para} and the conditions in Theorem \ref{thm:tensor} hold.
	Assume $\max_{j \in \mc{J}}|e_j-s_j+1|\leq c_1 \delta B$ for an absolute constant $c_1>0$, and the number of random intervals $J$ is large enough that  $J\geq 9\log (B\delta^{-1})/\delta^2$.
	Let   ${\epsilon}={k^2\nu}/{\underline{\lambda}^{2}}
	\leq {k^2\log(Np)}/{\underline{\lambda}^{2}}$.
	Suppose there exists a positive constant $c_2$ such that the CUSUM threshold satisfies $\sqrt{\epsilon} < c_2 \tau   < \sqrt{\delta B}\underline{\lambda}/k$.
	Then, with probability  at least $1- 3 N^{3-c_3}p^{2-c_3}-\delta^{-1}\exp(-\delta ^2 J/9)$ for some constant $c_3>3$, we have  $\hat{Q}={Q}$  and $\max_{q=1,\ldots,\hat{Q}}|\hat{u}_q-{u}_q|< c_4 \epsilon$ with some constant  $c_4>0$.
\end{theorem}

Note that the error is linear in $k^2$, which is of the same order of $k_0^2$ with $k\gtrsim k_0$.
Again, it emphasizes the advantage of accounting for the sparsity.
\cite{wang2021optimal} established the minimax lower bounds of the localization error for the problem of covariance change point for independent high dimensional sub-Gaussian data.
For our result, the localization error on the original scale is bounded by $\epsilon L={k^2 \log(Np)R}/{\underline{\lambda}^2}$
and we comment that this matches the minimax rate up to $\log(Np)R$, where $R$ is due to the existence of autocovariance matrices of order $-R$ to $R$ in (\ref{eq:fhat}).
As for the number of random intervals $J$, Theorem \ref{thm:cp} provides the choice $J\geq 9\log (B\delta^{-1})/\delta^2$. In the simple case that $\delta$ is of constant order, i.e., there are finite number of change points, this results in $J \sim log (B)$, which is low computational complexity. See also the discussion of computation complexity of the wild binary segmentation methods in \cite{fryzlewicz2014wild}.

\section{Empirical issues} \label{subsec:para}
We discuss in this section some aspects about the  implementation of Algorithm \ref{alg:multi}, including hyper-parameters choices and practical techniques, based on our limited experience gained from the simulation experiments reported below.

Our algorithm requires the input of the block length $L$ and the frequency set $\Omega$, both of which could be set based on the user's preference. 
For example, certain neuroscience data, such as the one in Section \ref{sec:app2}, may be segmented into uniform time intervals based on the sampling rate and analyzed within a biologically meaningful frequency range.
If no specific preference is indicated, we empirically use $L$ between 50 and 100, and $\Omega$ as a moderate number (typically 15-20) of evenly spaced frequencies within $[0,\pi]$.

We provide choices for other hyper-parameters.
Since $R \geq (L/\log(Np))^{1/3}$ by Assumption \ref{ass:para} (iii), we find $R=\lfloor L^{1/3} \rfloor \lor 1$ works well for our change point detection.
For Algorithms \ref{alg:tensor} and \ref{alg:multi},  we specify  a trimming distance $\nu\asymp{R \log(Np)}/{L} \geq (\log(Np)/{L})^{2/3}$
for technical reasons.
In practice, we set $\nu=0$ in Algorithm \ref{alg:tensor} to exploit full information in finding the projector, and $\nu=\lfloor (B\log(Np))^{2/3}/15\rfloor \lor 1$ in Algorithm \ref{alg:multi}.
Also, to avoid spurious change point detection, we require $\mf{C}_{s_j,e_j}(\hat{\mc{F}})$ to  stay positive in a small section around the detected change point.
Specifically, let $\hat{u}^{[j]}=\arg\max_u \mf{C}_{s_j,u,e_j}(\hat{\mc{F}})$ within the set $\{u: \min(u-s_j,e_j-u)>\nu, \mf{C}_{s_j,b,e_j}(\hat{\mc{F}})>0, |b-u| <\nu/4\}$.

As for the sparsity $k$,
we use a  heuristic approach to determine an upper bound of the number of series components sustaining at least one change.
We do this by 
implementing Algorithm \ref{alg:multi} with  each univariate component time series of $\mbf{X}$, where 
projection is not needed and hence is fast.
Then, we set $k$ to be the number of such time series found to have some change points.
The series-by-series approach tends to find spurious change points for stationary time series, making it non-competitive as a standalone method but providing a reasonable choice of $k$ for our proposed method since  Theorem \ref{thm:cp} requires $k$ to be greater  than $k_0$.
We recommend the wild interval number $J=2B$.
In the simulation studies, we found that even setting $J=30$ yielded satisfactory results.
In implementing Algorithm \ref{alg:multi} for $\mbf{X}$, we determine the threshold $\tau$ to be frequency-specific as $\tau(\omega)$ for $\omega \in \Omega$
via bootstrap as follows:   Generate bootstrap samples of $\hat{\mc{F}}(\omega)$ by re-sampling with replacement the block indexes, selecting only blocks whose spectral norm (averaged across frequencies) is below the 90th percentile to enhance robustness against potential outlier effects.
For each such bootstrap sample,  
run Algorithm \ref{alg:single} with $s=1$ and $e=B$, and calculate the value
$\max_{s \leq u \leq e}|\hat{\mc{T}}^*_{s,u,e}(\omega)|/\hat{\sigma}_{s,e}(\omega)$.
Take an upper quantile (97.5\%) of these  bootstrap values as $\tau(\omega)$.

To further enhance robustness against outliers and heavy tailed data,
a technique is to  apply a nonparametric  transformation to make the data marginally normal, via applying an instantaneous normal quantile transformation  co-ordinate by co-ordinate.
Specifically, let $F_N$ be the empirical marginal cumulative distribution function (CDF) of a particular component, and  $\Phi$ the CDF of the standard normal distribution.
Then the normal quantile tranformation is effected by the function $\Phi^{-1}\circ \{F_N-1/(2N)\}$.
For normally distributed data, the transformation is akin to  standardization.
We show the benefit of this technique for heavy-tailed data in Supplementary \ref{sec:simudist} and apply it in the real applications.

Our method achieves theoretically reliable detection of change points on the block scale, which is valuable for high-resolution time series such as neuroscience data, but can be improved for real applications where a finer resolution is desirable.
While one approach is to run our algorithm with shifting block starting positions, this increases computational cost.
Instead, we introduce an empirical local refinement step after block detection.
To account for the loss by the discretized block structure, we consider the neighbouring blocks  around each estimated location as potential finer candidates and scan them by a sliding window approach.
For each estimated location $\hat{u}$ from Algorithm \ref{alg:multi}, we retain the according projection vectors $\hat{\bm{\gamma}}(\omega)=\hat{\bm{\gamma}}^{[j_0]}(\omega)$ with $j_0$ from Line 11 in the algorithm for each frequency $\omega \in \Omega$.
For each candidate locations within the interval $n \in [(\hat{u}-1)L,(\hat{u}+1)L]$, we estimate the spectral density matrix in the left window $(n-L',n]$ as $\hat{\mc{F}}_n^-(\omega)$, and that in the right window $(n,n+L']$ as $\hat{\mc{F}}_n^+(\omega)$ with a window size $L'$.
Setting $L'=4L$ performs well in simulations.
Then, the refined change point is determined by maximizing the projected spectral contrast:
$\hat{u}'=\arg\max_{(\hat{u}-1)L\leq n \leq (\hat{u}+1)L}\max_{\omega \in \Omega}|\hat{\bm{\gamma}}(\omega)^{\top}(\hat{\mc{F}}_n^-(\omega)-\hat{\mc{F}}_n^+(\omega))\hat{\bm{\gamma}}(\omega)|$. 
This refinement step controls computational cost by leveraging the precomputed projection vectors and a simple contrast measure.
We test the empirical performance of our algorithm with this refinement step in Supplementary Section \ref{sec:simurefine}, and apply it in the change point estimation problem for the stock data in Section \ref{sec:app1}.

\section{Empirical performance}\label{sec:simulation}

\subsection{Simulation scenarios}\label{sec:dgp}

The main objectives of this simulation section are:
(i) to evaluate the performance of Algorithm \ref{alg:tensor},
(ii) to evaluate the performance of Algorithm \ref{alg:multi},
and (iii) to show the benefit of exploiting the sparsity structure.

We consider three data generating processes (DGPs) for verifying these three objectives.
Recalling that $\mbb{S}$ is defined as the CP series index set of cardinality $k_0$, which will be specified later.
\begin{align*}
\textbf{DGP1 (factor):}\qquad	\mbf{X}_n=\left\{
	\begin{matrix}
		&\mbf{Z}_{n}, & 0< n/N\leq v_1,\\
		&\mbf{Z}_{n}+\bm{\gamma} W_n, & v_1< n/N\leq 1,\\
	\end{matrix}
	\right.
\end{align*}
where $\bm{\gamma}$ is a unit $p$ vector with $\bm{\gamma}_i=1/\sqrt{k_0}$ for $i \in \mbb{S}$ and 0 otherwise, and $W_n= {\epsilon}_n +\phi {\epsilon}_{n-1}$ is a univariate MA(1) process with i.i.d.  normal errors ${\epsilon}_n \sim \mc{N}(0,\sigma^2)$ independent of $\mbf{Z}_{n}$.
$\mbf{Z}_n$ follows a  VMA(1) model with the VMA coefficient matrix $\mrm{diag}(0.6\times \mbf{1}_p)$, and i.i.d. errors from $\mc{N}({\bm{0}_p,\mbf{\Sigma}_{p \times p}})$ with $\mbf{\Sigma}_{ii}=2$ and $\mbf{\Sigma}_{ij}=0.2$ for $i \neq j$, where $\mbf{1}_p$ is the $p$-length vector of all 1s.
\begin{align*}
	\textbf{DGP2 (VMA):}\qquad	\mbf{X}_n=\left\{
	\begin{matrix}
		&\bm{\epsilon}_n+\mbf{\Phi}^{(1)} \bm{\epsilon}_{n-1}, & v_{q-1}< n/N \leq v_q \text{ for odd }q,\\
		&\bm{\epsilon}_n+\mbf{\Phi}^{(2)} \bm{\epsilon}_{n-1}, &  v_{q-1}< n/N \leq v_q \text{ for even }q,\\
	\end{matrix}
	\right.
\end{align*}
where  $\mbf{\Phi}^{(1)}=\mrm{diag}(0.6\times \mbf{1}_p)$, and  $\mbf{\Phi}^{(2)}=\mbf{\Phi}^{(1)}$ except that $\mbf{\Phi}^{(2)}_{ii}=-0.6$ for $i \in \mbb{S}$.
\begin{align*}
	\textbf{DGP3 (VAR):}	\mbf{X}_n=\left\{
	\begin{matrix}
		&\bm{\epsilon}_n+\mbf{\Psi}_1 \mbf{X}_{n-1}+\mbf{\Psi}_2^{(1)} \mbf{X}_{n-2}, & v_{q-1}< n/N \leq v_q \text{ for odd }q\\
		&\bm{\epsilon}_n+\mbf{\Psi}_1 \mbf{X}_{n-1}+\mbf{\Psi}_2^{(2)} \mbf{X}_{n-2}, &  v_{q-1}< n/N \leq v_q \text{ for even }q,\\
	\end{matrix}
	\right.
\end{align*}
where $\mbf{\Psi}_1=\mrm{diag}(0.1 \times \mbf{1}_p)$,  $\mbf{\Psi}_2^{(1)}=\mrm{diag}(0.4\times \mbf{1}_p)$, and $\mbf{\Psi}_2^{(2)}=\mbf{\Psi}_2^{(1)}$ except that $\mbf{\Psi}_{2,ii}^{(2)}=-0.7$ for $i \in \mbb{S}$.
For both DGP 2 and 3, on all segments, the innovations $\{\bm{\epsilon}_n\}$ are i.i.d. $\mc{N}({0_p,\mbf{\Sigma}_{p \times p}})$ with $\mbf{\Sigma}_{ii}=1$, and $\mbf{\Sigma}_{ij}=0.2$ for $i \neq j$.

We comment on the design of these DGPs.
DGP 1 follows  (\ref{eq:linear}), and hence $\mbf{g}(\omega)$ defined in (\ref{eq:g}) equals the spectrum of $\bm{\gamma} W_n$, with the eigenvalue (spectral norm) $\lambda_1(\omega)$ being the spectral density of $W_n$, see  Figure \ref{fig:proj}.
And it exactly satisfies the sparsity condition in Assumption \ref{ass:factor} with the vector $\gamma$ of cardinality $k_0$. 
This DGP facilitates straightforward manipulation of the spectral norm and the sparse projection, and hence will be used to validate objectives (i) and (iii).
For simplicity, we consider the   single-change-point setting here.

DGPs 2 and 3 consider commonly-used VMA and VAR models with coefficients changes in the multiple-change-points setting, respectively. Note that, given the contemporaneous correlation among elements of $\bm{\epsilon}_n$, the structure break of the spectrum also exists in components outside $\mbb{S}$. Hence, these two DGPs belong to the   ``approximately" sparse case, which allow us to assess the generality of our methodology.
What's more, for DGP2, when $\mbb{S}$ contains all the components, the two sub-models with $\mbf{\Phi}^{(1)}$ and $\mbf{\Phi}^{(2)}$ are not distinguishable in terms of the covariance matrix, indicating the necessity and importance of our method for exploiting the spectrum structure.

We contrast the empirical performances of the proposed Spectral Change Point method (SCP) with several state-of-art methods including: SBS-LSW \citep{cho2015multiplechangepoint}, which  uses the sparsified binary segmentation with the locally stationary wavelet periodogram; the factor-adjusted vector autoregressive method  \citep{cho2024highdimensional}, which detects structure break in both the factor driven component (FVAR-c) and the idiosyncratic VAR process (FVAR-i); the FreSpeD method  \citep{schroder2019fresped}, which detects change points in each auto-spectra, followed by a cluster postprocessing procedure.
We also include the series-by-series method as a benchmark, which implements Algorithm \ref{alg:multi} component-wise and aggregates the found change points via clustering.
See Supplementary Section  \ref{sec:auxsimulation} for further implementation  details of these methods.
For all  simulation results, each  experiment was replicated 500 times.

\subsection{Simulation results}\label{subsec:simures}
\begin{figure}
	\centering
	\includegraphics[width=0.9\textwidth]{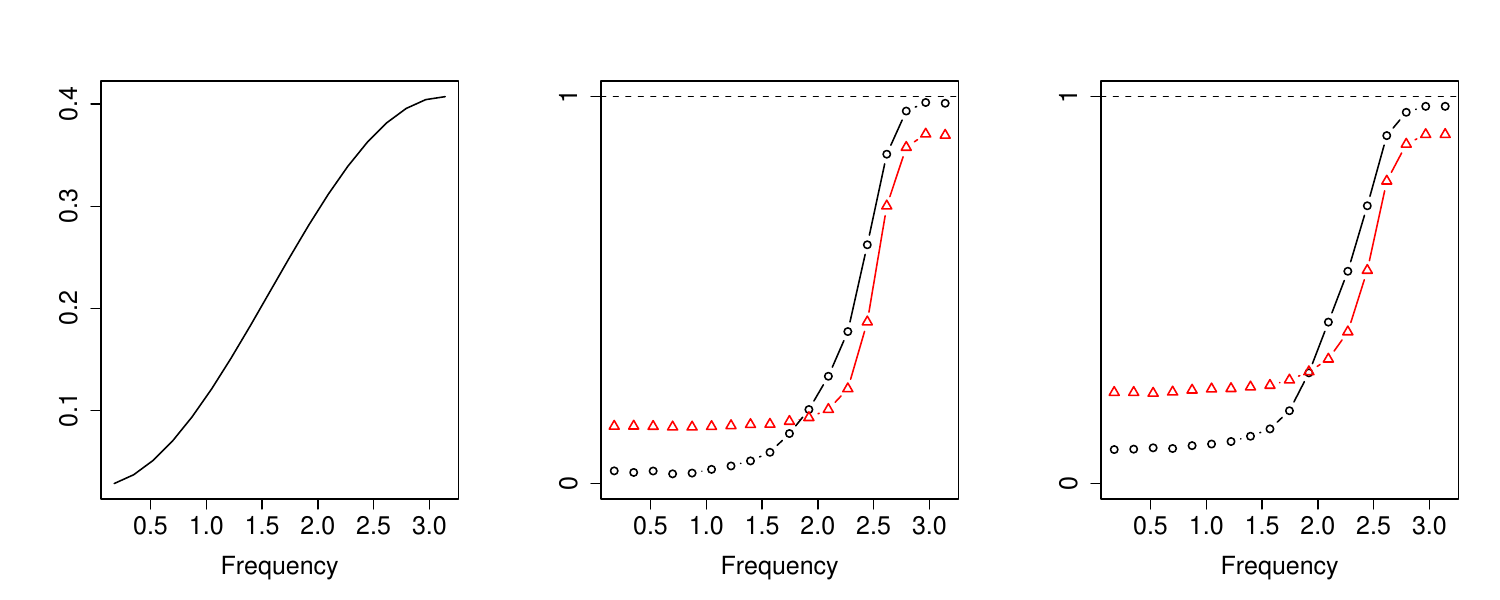}
	\caption{Left: $\lambda_1(\omega)$ versus $\omega$; $|\bm{\gamma}(\omega)^\top \hat{\bm{\gamma}}_{1,B}(\omega)|$ versus $\omega$ for $k_0=3$ (Middle) and $k_0=8$ (Right). $\circ: k=k_0$; 
		$\textcolor{red}{\vartriangle}:k=p$.}
	\label{fig:proj}
\end{figure}

For DGP1, we consider $p=80$, $N=6000$, $L=75$ and $v_1=0.5$.
Two sparsity levels are considered with $k_0  \in \{3,\lfloor\sqrt{p}\rfloor\}$, i.e., $\{3,8\}$, and $\mbb{S}$ is the set of the CP series that are evenly distributed across the $p$ dimensions. 
We first examine the performance of Algorithm \ref{alg:tensor} by evaluating $|\bm{\gamma}(\omega)^\top \hat{\bm{\gamma}}_{1,B}(\omega)|$.
For $W_n$ in DGP1, we set $\phi=-0.6$ and $\sigma^2=0.6$.
Figure \ref{fig:proj} plots the mean absolute inner product against frequency.
Besides $k=k_0$, we also consider $k=p$ as a benchmark.
Figure \ref{fig:proj} shows that the inner product curves bear resemblance to the signal magnitude $\lambda_1(\omega)$.
Over the  high frequency region when the signal is strong, the mean absolute inner product is generally close to 1 for $k=k_0$, indicating almost perfect alignment with its true counterpart $\bm{\gamma}$ in DGP1.
The inferior performance with $k=p$ shows the benefit of exploring the sparsity structure.

We also check the performance of change point detection for DGP1.
Specifically, for $W_n$ in DGP1, let $\phi=-0.3$ and $\sigma^2=0.6$ or $\sigma^2=5$, respectively representing weak  and strong  signals.
Again, we consider $k_0  \in 
\{3,8\}$. 
In Table \ref{tab:factor}, we report the simulation results of the estimated change point numbers and locations. 
We show the frequencies of the estimated numbers being less than, equal to or greater  than the true value.
As for the change point location, we use the adjusted Rand index (ARI) of the estimated
partition of the time points into stationary segments against the true partition to measure the performance, since change point estimation  may be viewed as a special case of classification.

\begin{table}[htbp]

\caption{Simulation results for DGP1}
\label{tab:factor}
\centering
\footnotesize
\resizebox{0.8\linewidth}{!}{
\begin{tabular}[t]{lr>{}rr>{}rr>{}rr>{}r}
\toprule
\multicolumn{1}{c}{ } & \multicolumn{4}{c}{$\sigma^2=0.6,  k_0=3$} & \multicolumn{4}{c}{$\sigma^2=5, k_0=8$} \\
\cmidrule(l{3pt}r{3pt}){2-5} \cmidrule(l{3pt}r{3pt}){6-9}
  & $\hat{q}<q$ & $\hat{q}=q$ & $\hat{q}>q$ & ARI & $\hat{q}<q$ & $\hat{q}=q$ & $\hat{q}>q$ & ARI\\
\midrule
SCP & 0.00 & \cellcolor{gray!6}{0.91} & 0.09 & \cellcolor{gray!6}{0.95} & 0.00 & \cellcolor{gray!6}{0.99} & 0.01 & \cellcolor{gray!6}{0.99}\\
SBS-LSW & 0.13 & \cellcolor{gray!6}{0.80} & 0.07 & \cellcolor{gray!6}{0.73} & 0.00 & \cellcolor{gray!6}{0.90} & 0.10 & \cellcolor{gray!6}{0.98}\\
Series-by-series & 0.00 & \cellcolor{gray!6}{0.06} & 0.94 & \cellcolor{gray!6}{0.65} & 0.00 & \cellcolor{gray!6}{0.14} & 0.86 & \cellcolor{gray!6}{0.70}\\
FVAR-c & 1.00 & \cellcolor{gray!6}{0.00} & 0.00 & \cellcolor{gray!6}{0.00} & 0.04 & \cellcolor{gray!6}{0.96} & 0.00 & \cellcolor{gray!6}{0.94}\\
FVAR-i & 1.00 & \cellcolor{gray!6}{0.00} & 0.00 & \cellcolor{gray!6}{0.00} & 1.00 & \cellcolor{gray!6}{0.00} & 0.00 & \cellcolor{gray!6}{0.00}\\
FreSpeD & 0.32 & \cellcolor{gray!6}{0.42} & 0.26 & \cellcolor{gray!6}{0.34} & 0.00 & \cellcolor{gray!6}{0.03} & 0.97 & \cellcolor{gray!6}{0.70}\\
SCP ($k=k_0$) & 0.00 & \cellcolor{gray!6}{0.92} & 0.08 & \cellcolor{gray!6}{0.95} & 0.00 & \cellcolor{gray!6}{0.97} & 0.03 & \cellcolor{gray!6}{0.99}\\
SCP ($k=p$) & 0.31 & \cellcolor{gray!6}{0.68} & 0.01 & \cellcolor{gray!6}{0.59} & 0.00 & \cellcolor{gray!6}{1.00} & 0.00 & \cellcolor{gray!6}{1.00}\\
\bottomrule
\end{tabular}}
\end{table}

Table \ref{tab:factor} shows that  our method is very competitive in terms of accuracy regarding the number and location of the change points.
To compare, the SBS-LSW and FVAR method behave well in  the strong signal case ($\sigma^2=5, k_0=8$), but have a tendency to miss the change point in the weak signal case ($\sigma^2=0.6, k_0=3$).
As for the series-by-series and the FreSpeD approaches,
both employing a procedure where marginal detection is conducted first followed by clustering, they have a tendency to overestimate change points. 
Particularly, these methods may introduce spurious change points whenever any component detects one, rendering it less effective in high dimensions.
In contrast, with information integration over the entire  matrix, our method is more stable against spurious change points.
Our SCP method sets the sparsity parameter $k$ via  the data-driven approach elaborated in Section \ref{subsec:para}.
To compare, we have also examined the performance when $k$ is set to the true $k_0$ or $p$.
It turns out that our method performs as well as the one with true $k_0$.
However, using $k=p$ may result in significant  performance loss, particularly when the signal is weak, possibly due to the aggregation of noises. 
In this more challenging cases with weaker signals, capturing information from the signals not affected by noise becomes crucial to improve detection accuracy.
This once again underscores the advantages of exploring the sparsity structure inherent in the data.

We consider multiple change points for DGP2 and DGP3 with $q=4$ equally spaced change points.
For $p=80$, we consider different sparsity $k_0$ from $\{3,\lfloor\sqrt{p}\rfloor,{p}/2,p\}$, i.e., $\{3,8,40,80\}$.
The elements of the CP series index set $\mbb{S}$ are  evenly distributed across the $p$ dimensions. 
We consider $N=6000, 12000$ with $L=75$, and we only show the simulation results for DGP2 with $N=6000$ and DGP3 with $N=12000$ in Tables \ref{tab:vma} and \ref{tab:var}; see the other two tables in the Supplementary Section \ref{sec:auxsimulation}.

\begin{table}[htbp]
\caption{Simulation results for DGP2 with $N=6000$}
\label{tab:vma}
\centering
\resizebox{0.9\linewidth}{!}{
\begin{tabular}[t]{rrr>{}rr>{}rr>{}rr>{}rr>{}rr>{}r}
\toprule
$N$ & $k_0$ & $\hat{q}<q$ & $\hat{q}=q$ & $\hat{q}>q$ & ARI & $\hat{q}<q$ & $\hat{q}=q$ & $\hat{q}>q$ & ARI & $\hat{q}<q$ & $\hat{q}=q$ & $\hat{q}>q$ & ARI\\
\midrule
\multicolumn{1}{c}{ } & \multicolumn{1}{c}{ } & \multicolumn{4}{c}{SCP} & \multicolumn{4}{c}{SBS-LSW} & \multicolumn{4}{c}{Series-by-series} \\
\cmidrule(l{3pt}r{3pt}){3-6} \cmidrule(l{3pt}r{3pt}){7-10} \cmidrule(l{3pt}r{3pt}){11-14}
& 3 & 0.02 & \cellcolor{gray!6}{0.98} & 0.00 & \cellcolor{gray!6}{0.98} & 0 & \cellcolor{gray!6}{0.82} & 0.18 & \cellcolor{gray!6}{0.98} & 0 & \cellcolor{gray!6}{0.19} & 0.81 & \cellcolor{gray!6}{0.90}\\

 & 8 & 0.01 & \cellcolor{gray!6}{0.99} & 0.00 & \cellcolor{gray!6}{0.99} & 0 & \cellcolor{gray!6}{0.76} & 0.24 & \cellcolor{gray!6}{0.98} & 0 & \cellcolor{gray!6}{0.21} & 0.79 & \cellcolor{gray!6}{0.90}\\

 & 40 & 0.00 & \cellcolor{gray!6}{1.00} & 0.00 & \cellcolor{gray!6}{1.00} & 0 & \cellcolor{gray!6}{0.31} & 0.69 & \cellcolor{gray!6}{0.95} & 0 & \cellcolor{gray!6}{0.42} & 0.58 & \cellcolor{gray!6}{0.91}\\

 & 80 & 0.00 & \cellcolor{gray!6}{1.00} & 0.00 & \cellcolor{gray!6}{1.00} & 0 & \cellcolor{gray!6}{0.11} & 0.89 & \cellcolor{gray!6}{0.92} & 0 & \cellcolor{gray!6}{0.92} & 0.08 & \cellcolor{gray!6}{0.95}\\
 \multicolumn{2}{c}{ } & \multicolumn{4}{c}{FVAR-c} & \multicolumn{4}{c}{FVAR-i} & \multicolumn{4}{c}{FreSpeD} \\
 \cmidrule(l{3pt}r{3pt}){3-6} \cmidrule(l{3pt}r{3pt}){7-10} \cmidrule(l{3pt}r{3pt}){11-14}
 \addlinespace[0.1em]
 & 3 & 1.00 & \cellcolor{gray!6}{0.00} & 0.00 & \cellcolor{gray!6}{0.00} & 0 & \cellcolor{gray!6}{1.00} & 0.00 & \cellcolor{gray!6}{0.98} & 0 & \cellcolor{gray!6}{0.08} & 0.92 & \cellcolor{gray!6}{0.89}\\

 & 8 & 0.93 & \cellcolor{gray!6}{0.07} & 0.00 & \cellcolor{gray!6}{0.33} & 0 & \cellcolor{gray!6}{1.00} & 0.00 & \cellcolor{gray!6}{0.98} & 0 & \cellcolor{gray!6}{0.01} & 0.99 & \cellcolor{gray!6}{0.80}\\

 & 40 & 0.00 & \cellcolor{gray!6}{0.92} & 0.08 & \cellcolor{gray!6}{0.97} & 0 & \cellcolor{gray!6}{1.00} & 0.00 & \cellcolor{gray!6}{0.98} & 0 & \cellcolor{gray!6}{0.00} & 1.00 & \cellcolor{gray!6}{0.64}\\

\multirow{-8}{*}{\raggedleft\arraybackslash 6000} & 80 & 0.00 & \cellcolor{gray!6}{0.72} & 0.28 & \cellcolor{gray!6}{0.95} & 0 & \cellcolor{gray!6}{1.00} & 0.00 & \cellcolor{gray!6}{0.98} & 0 & \cellcolor{gray!6}{0.00} & 1.00 & \cellcolor{gray!6}{0.66}\\
\bottomrule
\end{tabular}}
\end{table}

\begin{table}[htbp]
\caption{Simulation results for DGP3 with $N=12000$}
\label{tab:var}
\centering
\resizebox{0.9\linewidth}{!}{
\begin{tabular}[t]{rrr>{}rr>{}rr>{}rr>{}rr>{}rr>{}r}
\toprule
$N$ & $k_0$ & $\hat{q}<q$ & $\hat{q}=q$ & $\hat{q}>q$ & ARI & $\hat{q}<q$ & $\hat{q}=q$ & $\hat{q}>q$ & ARI & $\hat{q}<q$ & $\hat{q}=q$ & $\hat{q}>q$ & ARI\\
\midrule
\multicolumn{1}{c}{ } & \multicolumn{1}{c}{ } & \multicolumn{4}{c}{SCP} & \multicolumn{4}{c}{SBS-LSW} & \multicolumn{4}{c}{Series-by-series} \\
\cmidrule(l{3pt}r{3pt}){3-6} \cmidrule(l{3pt}r{3pt}){7-10} \cmidrule(l{3pt}r{3pt}){11-14}
 & 3 & 0.0 & \cellcolor{gray!6}{1.00} & 0.00 & \cellcolor{gray!6}{0.99} & 0 & \cellcolor{gray!6}{0.66} & 0.34 & \cellcolor{gray!6}{0.96} & 0 & \cellcolor{gray!6}{0.29} & 0.71 & \cellcolor{gray!6}{0.92}\\

 & 8 & 0.0 & \cellcolor{gray!6}{1.00} & 0.00 & \cellcolor{gray!6}{0.99} & 0 & \cellcolor{gray!6}{0.45} & 0.55 & \cellcolor{gray!6}{0.96} & 0 & \cellcolor{gray!6}{0.32} & 0.68 & \cellcolor{gray!6}{0.92}\\

 & 40 & 0.0 & \cellcolor{gray!6}{1.00} & 0.00 & \cellcolor{gray!6}{0.99} & 0 & \cellcolor{gray!6}{0.02} & 0.98 & \cellcolor{gray!6}{0.88} & 0 & \cellcolor{gray!6}{0.54} & 0.46 & \cellcolor{gray!6}{0.94}\\

 & 80 & 0.0 & \cellcolor{gray!6}{0.99} & 0.01 & \cellcolor{gray!6}{0.99} & 0 & \cellcolor{gray!6}{0.00} & 1.00 & \cellcolor{gray!6}{0.84} & 0 & \cellcolor{gray!6}{0.93} & 0.07 & \cellcolor{gray!6}{0.97}\\
 \multicolumn{2}{c}{ } & \multicolumn{4}{c}{FVAR-c} & \multicolumn{4}{c}{FVAR-i} & \multicolumn{4}{c}{FreSpeD} \\
\cmidrule(l{3pt}r{3pt}){3-6} \cmidrule(l{3pt}r{3pt}){7-10} \cmidrule(l{3pt}r{3pt}){11-14}
\addlinespace[0.1em]
& 3 & 1.0 & \cellcolor{gray!6}{0.00} & 0.00 & \cellcolor{gray!6}{0.03} & 1 & \cellcolor{gray!6}{0.00} & 0.00 & \cellcolor{gray!6}{0.00} & 0 & \cellcolor{gray!6}{0.58} & 0.42 & \cellcolor{gray!6}{0.97}\\

 & 8 & 0.1 & \cellcolor{gray!6}{0.90} & 0.00 & \cellcolor{gray!6}{0.95} & 1 & \cellcolor{gray!6}{0.00} & 0.00 & \cellcolor{gray!6}{0.00} & 0 & \cellcolor{gray!6}{0.28} & 0.72 & \cellcolor{gray!6}{0.94}\\

 & 40 & 0.0 & \cellcolor{gray!6}{0.69} & 0.31 & \cellcolor{gray!6}{0.96} & 1 & \cellcolor{gray!6}{0.00} & 0.00 & \cellcolor{gray!6}{0.00} & 0 & \cellcolor{gray!6}{0.00} & 1.00 & \cellcolor{gray!6}{0.83}\\

\multirow{-8}{*}{\raggedleft\arraybackslash 12000} & 80 & 0.0 & \cellcolor{gray!6}{0.71} & 0.29 & \cellcolor{gray!6}{0.96} & 1 & \cellcolor{gray!6}{0.00} & 0.00 & \cellcolor{gray!6}{0.01} & 0 & \cellcolor{gray!6}{0.00} & 1.00 & \cellcolor{gray!6}{0.78}\\
\bottomrule
\end{tabular}}
\end{table}

\begin{figure}[htbp]
\centering
\subfigure[]{
\begin{minipage}{3.5cm}
\centering
\includegraphics[width=\textwidth,height=\textwidth]{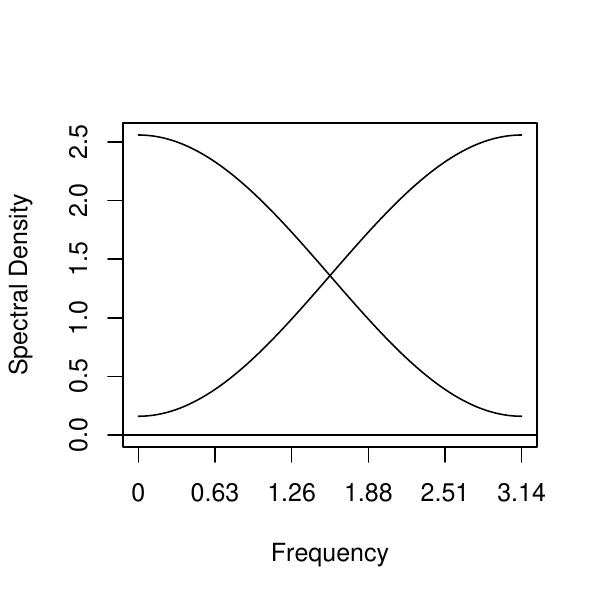}
\end{minipage}
}
\subfigure[]{
\begin{minipage}{3.5cm}
\centering
\includegraphics[width=\textwidth,height=\textwidth]{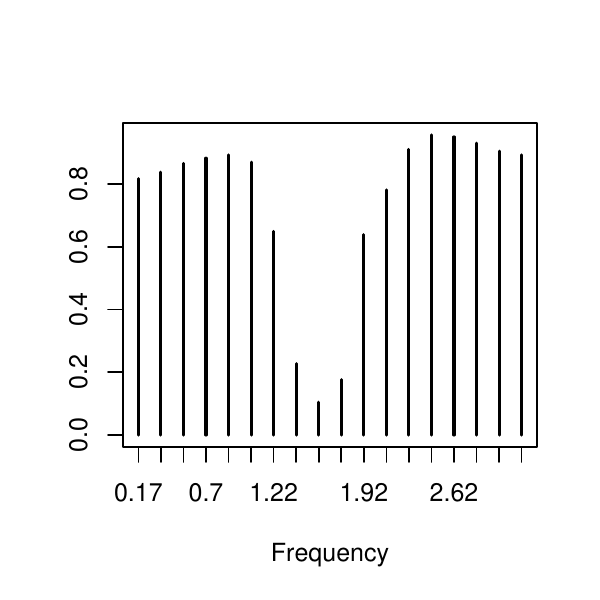}
\end{minipage}
}
\subfigure[]{
\begin{minipage}{3.5cm}
\centering
\includegraphics[width=\textwidth,height=\textwidth]{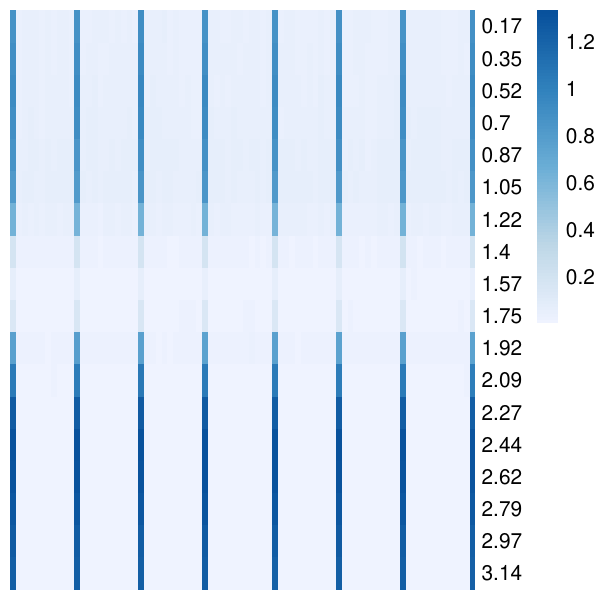}
\end{minipage}
}
\\
\subfigure[]{
\begin{minipage}{3.5cm}
\centering
\includegraphics[width=\textwidth,height=\textwidth]{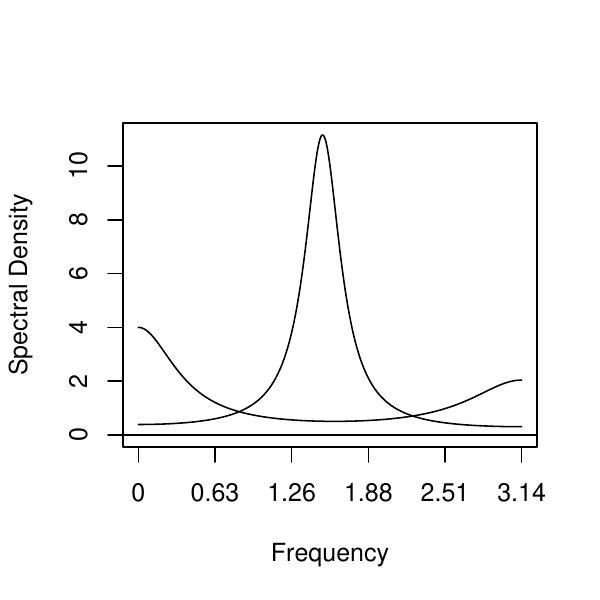}
\end{minipage}
}
\subfigure[]{
\begin{minipage}{3.5cm}
\centering
\includegraphics[width=\textwidth,height=\textwidth]{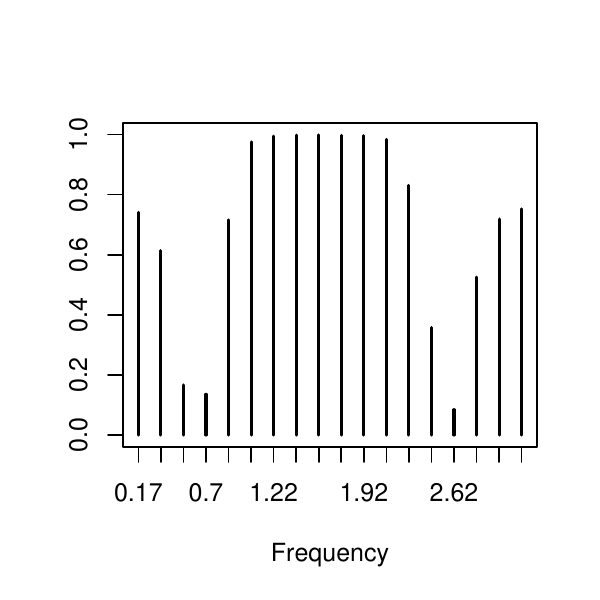}
\end{minipage}
}
\subfigure[]{
\begin{minipage}{3.5cm}
\centering
\includegraphics[width=\textwidth,height=\textwidth]{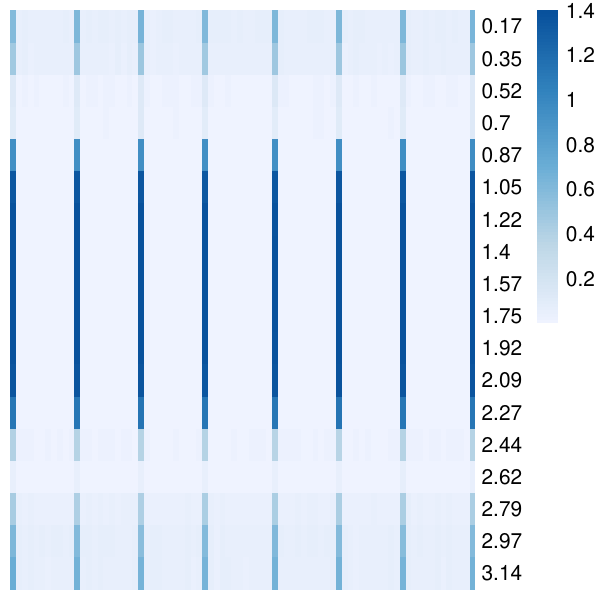}
\end{minipage}
}
\caption{Simulation results for DGP2 (Top row) and DGP3 (Bottom row). (a): Spectral densities of MA(1) with coefficient 0.6 and -0.6, 
respectively. 
(d): Spectral densities of AR(2) with coefficients 0.1, 0.4 and 0.1, -0.7, respectively.
(b), (e): Relative frequency of detecting a change point per frequency.
(c), (f): Averaged projection (in absolute value) per frequency (each row) and series (each column).
}
\label{fig:vmavar}
\end{figure}

Across experiments, the proposed  SCP can accurately estimate the number and location of changes across different scenarios.
And it adapts to different sparsity levels. 
The FVAR method is very competitive for DGP2, but not for DGP3, where the signal seems weaker.
We note that other methods may overestimate the change point numbers.
Next we consider the CP frequency and CP series.
Figure \ref{fig:vmavar} displays the results when $k_0=8$.
Clearly,  the shape of the empirical distribution of the CP frequencies bears resemblance to that of the difference in the spectral densities pre- and post-change.
And the projection correctly identifies the CP series.

In Supplementary Section \ref{sec:auxsimulation}, we report additional simulation results with different series lengths and dimensions. We also present the empirical performance of the proposed method against  non-normal  distributions.
Empirical performance of our algorithm with the refinement step introduced in Section \ref{subsec:para} is also examined.

\section{Applications}\label{sec:application}

\subsection{Analysis of the stock data}\label{sec:app1}
We illustrate the proposed method with the difference of log returns of the daily adjusted closing values for a subset of the Standard and Poor’s 100 (S\&P100) stocks, from November 1, 1999 to October 7, 2021, totalling $N = 5520$ observations.
Among the S\&P100 stocks, we focus our analysis on the $p = 79$ stocks with complete observations during this period.
The dataset is publicly available in the  $\mathtt{alphavantager}$ package of R.
%

\begin{figure}
	\includegraphics[width=0.23\textwidth,height=0.3\textwidth]{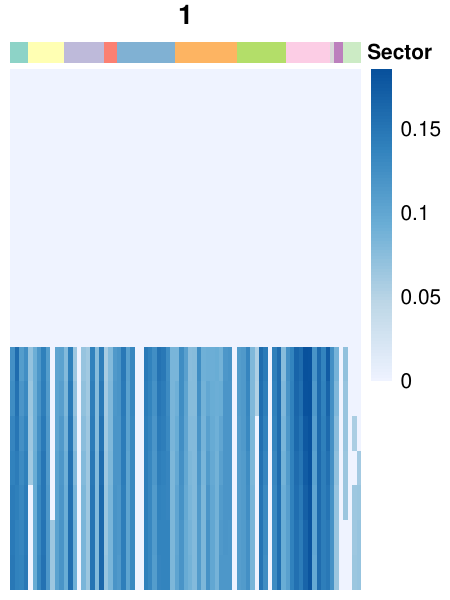}
	\includegraphics[width=0.23\textwidth,height=0.3\textwidth]{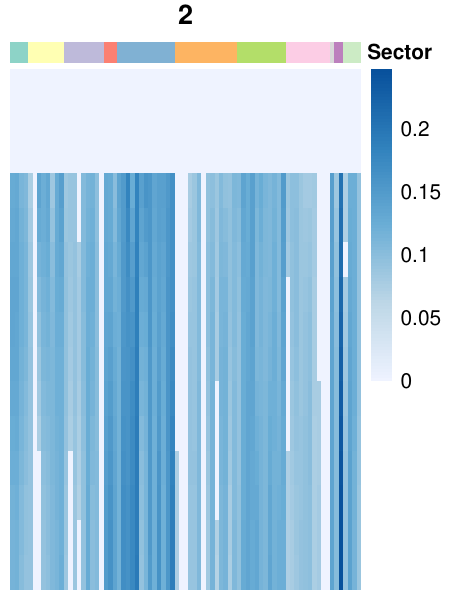}
	\includegraphics[width=0.23\textwidth,height=0.3\textwidth]{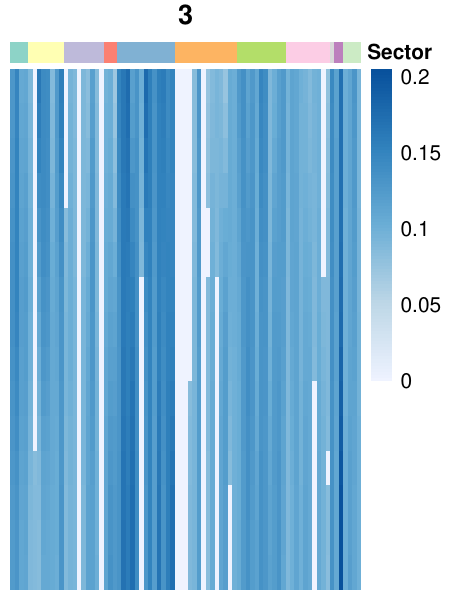}
	\includegraphics[width=0.29\textwidth,height=0.3\textwidth]{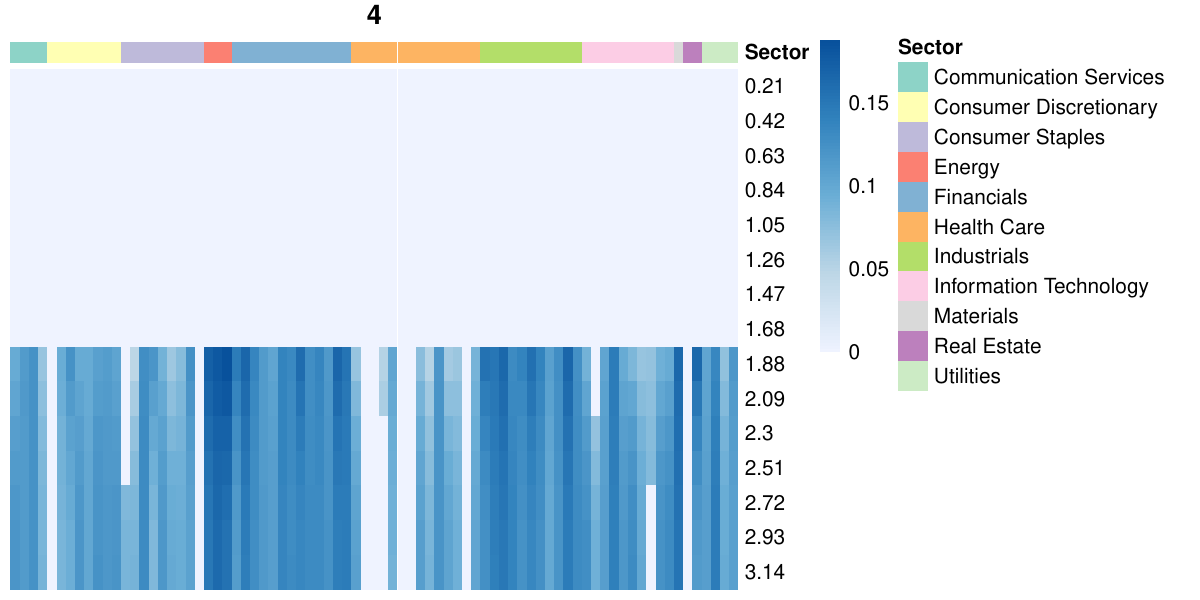}
	\caption{Projection per frequency (each row) and series (each column) for the four change points.
	}
	\label{fig:spcp}
\end{figure}

To apply Algorithm \ref{alg:multi}, we set the block lengths to be $L=60$, which is about three months.
We choose other hyper-parameters according to the guidance in Section \ref{subsec:para} and apply the normal quantile transformation and localized refinement step described therein.
Four change points are found, namely, 
at 2003-08-29, 2007-07-13, 2010-08-03 and 2018-09-18.
Figure \ref{fig:spcp} shows which series at which frequency underwent a change in their spectrum per change point, i.e., it displays the CP frequency set $\mbb{F}_q$ and the corresponding frequency-specific CP series set $\mbb{S}_q^{\omega}$ for $1 \leq q \leq 4$.

The four change points display quite different patterns.
At the 1st change point, the changes emerge in part of the frequencies (high frequencies), and most affected series (CP series) are stocks from the Information Technology sector.
This corresponds to the technology bubble burst, which heavily impacted IT companies.
The 2nd and 3rd change points show similarity in terms of the broad CP frequency and evenly-spread CP series, with the  Financial and Real Estate stocks having slightly greater weights.
These two change points correspond to the start and end of the subprime mortgage crisis and  global financial crisis.
For the 4th change point, the changes occurred over part of the frequencies, and the Energy sector stands out as CP series.
In 2019, there were significant fluctuations in oil prices due to various factors such as changes in global oil supply and demand dynamics and economic uncertainties. 
In this application, the CP series cast light on the processes underlying  the found structural breaks.

The change points we detected indicate structural breaks in the second-order structure of the data. Since stock data is commonly analyzed by economists and financial analysts using tools such as generalized dynamic factor models, our findings suggest that accounting for multiple change points in modeling could lead to a more accurate representation of its dynamic structure, and better capture shifts in market conditions and dependencies over time \citep{barigozzi2018simultaneous}. This insight can help guide future analyses, improving the interpretation and forecasting of financial data.

\subsection{Analysis of neonatal seizure}\label{sec:app2}
\begin{figure}
	\centering
	\includegraphics[width=0.3\textwidth,height=0.3\textwidth]{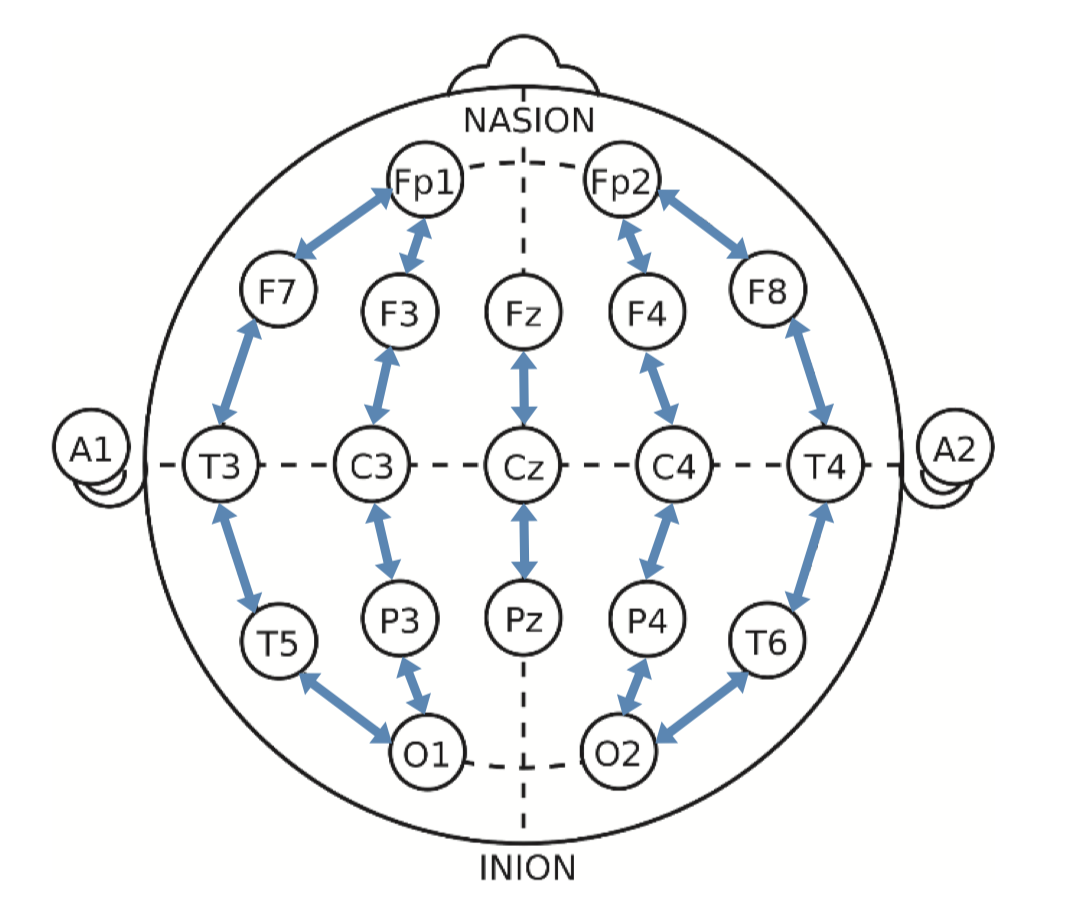}
	\includegraphics[width=0.6\textwidth,height=0.3\textwidth]{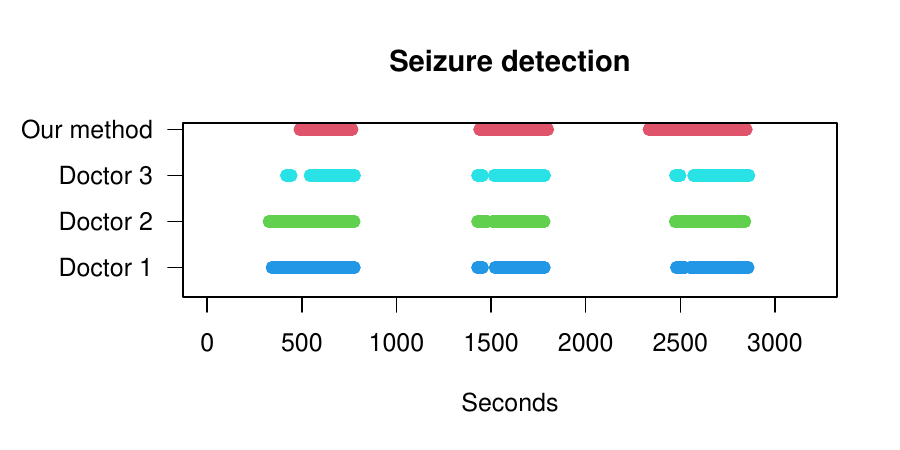}
	\caption{Left: the 19 electrodes position in \cite{stevenson2019dataset}. Right: the seizure periods detected by our method and annotated by three doctors.  
	}
	\label{fig:elec}
\end{figure}

Seizures are a common emergency in the neonatal intensive care unit (NICU) and pose a significant risk to the developing brain. 
Since most neonatal seizures lack obvious clinical manifestations, the gold standard for detecting seizures so far has been the visual analysis of long-duration electroencephalographic (EEG) recordings, However, interpreting neonatal EEGs is a time-consuming process that demands specialized expertise, which may not always be readily accessible.
The EEG data during neonatal seizure often manifest repetitive and periodic characteristics, making our spectral change point method a good fit to distinguish seizure periods from normal periods.
The EEG data is collected from a full-term neonate with severe bilateral asphyxia, recorded at the NICU of the Children’s Hospital, Helsinki University Central Hospital, Finland \citep{stevenson2019dataset}.
For recordings, 19 electrodes were placed according to the international 10-20 system, as shown in Figure \ref{fig:elec}, and the data is sampled at 256 Hz (256 observations per second) for 53.3 mins.
Hence, we have $N=819200$ observations at $p=19$ electrodes.
Preprocessing procedures include a notch filter at 50 Hz, which is the power line frequency, and a band-pass filter ranging from 0.5-30 Hz for its physiological relevance.
Along with the EEG recording, the data was annotated for the presence of seizure independently by three clinical experts.


In this analysis, we implement our proposed SCP method on the data to identify change points in the EEG, and explore the influence of seizure over frequencies.
Specifically, for the choice of $L$ and $\Omega$, we use 10s blocks, i.e. $L=2560$, and
we study on frequency 1-30 Hz with 1 Hz steps, including the delta bands (1-4 Hz), theta bands (4-8 Hz), alpha bands (8-13 Hz), and beta bands (13-30 Hz).
A total of six change points were identified, dividing the entire time period into seven stationary segments.
As shown in Figure \ref{fig:elec}, the separation aligns quite well with the doctors' opinions on seizure attack.
Specifically, the 2nd, 4th and 6th segments correspond to the three seizure periods annotated by the doctors. 
Although there may be some disagreement regarding the exact timing of seizures, we emphasize that this challenge is also present among doctors. 
Generally, our alignment is quite strong, and it serves as an expert tool to provide reliable and accurate seizure detection.

\begin{figure}
	\centering
	\includegraphics[width=0.7\textwidth,height=0.5\textwidth]{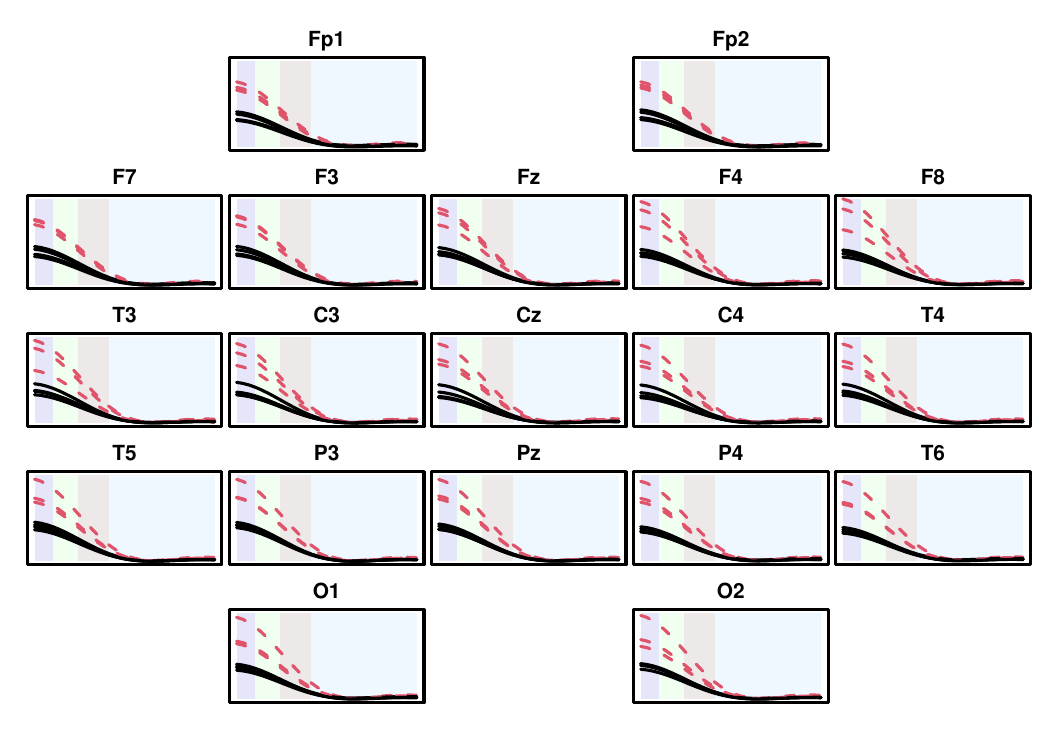}
	\caption{The mean of power over frequency on 7 segments separated by the estimated change point. Black solid curves for the 1st, 3rd, 5th and 7th segments (non-seizure periods), and red dash curves for 2nd, 4th and 6th segments(seizure periods). The x-axis range from 1-30 Hz, with the background color indicating four frequency bands. The plots are arranged according to the electrodes position in Figure \ref{fig:elec}.}
	\label{fig:power}
\end{figure}

In Figure \ref{fig:power}, the mean of the power over blocks in each of these seven segments is shown, and arranged by the  electrodes positions as in Figure \ref{fig:elec}.
The coherence between each pair are not shown due to the high dimension.
Each element in the matrix includes four black solid power curves over frequencies from 1 to 30 Hz during non-seizure periods and three red dash curves during the seizure periods. 
A clear separation is observed between the black and red curves for almost all elements.
Specifically, the black curves quite overlap, indicating a normal state.
A substantial increase in the 1-13 Hz range is observed in all power curves with possible different magnitudes, potentially corresponding to different seizure strength. 
The slow-wave frequency bands are  prominent in the neonatal brain, where increased activities might suggest a disruption in normal brain function.
These results not only demonstrate the utility of our spectral change point method in identifying breaks in brain network but also provide insights into the specific alterations within the brain network associated with neonatal seizures.
This serves as an initial step in exploring future topics on neonatal seizures, including the heterogeneous responses of multiple individuals, and the effects of anti-seizure medications in reducing changes.

\section{Discussion}\label{sec:discussion}
In this paper, we study the problem of change point detection within the general framework  of a high dimensional piecewise stationary process, via frequency domain analysis.
We accommodate both cross-sectional and temporal dependencies in the data as well as in the structural change.
By studying the general framework of changes in the spectral density  matrix, we encompass many standard parametric models, such as VAR and dynamic factor models.
The structure change is allowed to be series sparse and frequency specific, which facilitates practical analysis to discern its   origin and impact.
Specifically, the benefit of exploring the sparse structure is shown via both theory and empirical evidence.
We have shown that the proposed method, via projection based on a novel decomposition of the third-order CUSUM tensor, leads to stability and accuracy of the detected change points, as compared with existing methods. 

A number of interesting problems await future research.
There is potential for a more comprehensive exploration of frequency-specific information, particularly in fields like neuroscience, which constitutes an ongoing project for us.
Another topic is to detect change points with data  admitting other structures associated to change points, for instance, time series may be divided into groups whose  member series may likely experience a change together.  
Additionally, handling high-dimensional time series with both mixed spectra and structural breaks presents a challenging yet important direction for further study.

\section*{Acknowledgments 
}
The authors gratefully thank the anonymous referees, the associate editor and the editors for carefully reading
the paper and providing constructive comments that have improved the manuscript.

\section*{Conflicts of interest}
The authors do not declare any conflict of interest regarding this work.

\section*{Data Availability}
The data used in this manuscript is publicly available. R code is available from
\url{https://github.com/xinyuzhang15/SpecCp}.

\section*{Supplementary Material}
The \emph{Supplementary Materials} includes all proofs, more discussions as well as further numerical results. It is available online at Journal of the Royal Statistical Society: Series B.

\bibliographystyle{abbrvnat}%
\bibliography{library.bib}

@article{adak1998timedependent,
  title = {Time-Dependent Spectral Analysis of Nonstationary Time Series},
  author = {Adak, Sudeshna},
  year = {1998},
  journal = {Journal of the American Statistical Association},
  volume = {93},
  number = {444},
  pages = {1488--1501},
  doi = {10.1080/01621459.1998.10473808},
  langid = {english}
}

@inproceedings{allen2012sparse,
  title = {Sparse Higher-Order Principal Components Analysis},
  booktitle = {Artificial {{Intelligence}} and {{Statistics}}},
  author = {Allen, Genevera},
  year = {2012},
  pages = {27--36},
  publisher = {PMLR},
  langid = {english}
}

@article{aue2009break,
  title = {Break Detection in the Covariance Structure of Multivariate Time Series Models},
  author = {Aue, Alexander and H{\"o}rmann, Siegfried and Horv{\'a}th, Lajos and Reimherr, Matthew},
  year = {2009},
  journal = {The Annals of Statistics},
  volume = {37},
  number = {6B},
  eprint = {0911.3796},
  pages = {4046--4087},
  doi = {10.1214/09-AOS707},
  archiveprefix = {arXiv},
  langid = {english}
}

@article{barigozzi2018simultaneous,
  title = {Simultaneous Multiple Change-Point and Factor Analysis for High-Dimensional Time Series},
  author = {Barigozzi, Matteo and Cho, Haeran and Fryzlewicz, Piotr},
  year = {2018},
  journal = {Journal of Econometrics},
  volume = {206},
  number = {1},
  pages = {187--225},
  doi = {10.1016/j.jeconom.2018.05.003},
  langid = {english}
}

@book{brillinger2001time,
  title = {Time {{Series}}: {{Data Analysis}} and {{Theory}}},
  shorttitle = {Time {{Series}}},
  author = {Brillinger, David R.},
  year = {2001},
  publisher = {SIAM},
  googlebooks = {3DFJfgEW94gC},
  langid = {english}
}

@article{chan2014group,
  title = {Group {{LASSO}} for Structural Break Time Series},
  author = {Chan, Ngai Hang and Yau, Chun Yip and Zhang, Rong-Mao},
  year = {2014},
  journal = {Journal of the American Statistical Association},
  volume = {109},
  number = {506},
  pages = {590--599},
  doi = {10.1080/01621459.2013.866566},
  langid = {english}
}

@article{cho2012multiscale,
  title = {Multiscale and Multilevel Technique for Consistent Segmentation of Nonstationary Time Series},
  author = {Cho, Haeran and Fryzlewicz, Piotr},
  year = {2012},
  journal = {Statistica Sinica},
  volume = {22},
  number = {1},
  pages = {207--229},
  publisher = {JSTOR}
}

@article{cho2015multiplechangepoint,
  title = {Multiple-Change-Point Detection for High Dimensional Time Series via Sparsified Binary Segmentation},
  author = {Cho, Haeran and Fryzlewicz, Piotr},
  year = {2015},
  journal = {Journal of the Royal Statistical Society. Series B (Statistical Methodology)},
  volume = {77},
  number = {2},
  pages = {475--507},
  publisher = {[Royal Statistical Society, Wiley]}
}

@article{cho2024highdimensional,
  title = {High-Dimensional Time Series Segmentation via Factor-Adjusted Vector Autoregressive Modelling},
  author = {Cho, Haeran and Maeng, Hyeyoung and Eckley, Idris A. and Fearnhead, Paul},
  year = {2024},
  journal = {Journal of the American Statistical Association},
  volume = {119},
  number = {547},
  pages = {2038--2050},
  doi = {10.1080/01621459.2023.2240054},
  langid = {english}
}

@book{csorgo1997limit,
  title = {Limit Theorems in Change-Point Analysis},
  author = {Cs{\"o}rg{\"o}, M. and Horv{\'a}th, Lajos},
  year = {1997},
  series = {Wiley Series in Probability and Statistics},
  publisher = {Wiley},
  address = {Chichester ; New York},
  lccn = {QA276 .C87 1997}
}

@article{davis2006structural,
  title = {Structural Break Estimation for Nonstationary Time Series Models},
  author = {Davis, Richard A. and Lee, Thomas C. M. and {Rodriguez-Yam}, Gabriel A.},
  year = {2006},
  journal = {Journal of the American Statistical Association},
  volume = {101},
  number = {473},
  pages = {223--239},
  publisher = {[American Statistical Association, Taylor \& Francis, Ltd.]}
}

@article{enikeeva2019highdimensional,
  title = {High-Dimensional Change-Point Detection under Sparse Alternatives},
  author = {Enikeeva, Farida and Harchaoui, Zaid},
  year = {2019},
  journal = {The Annals of Statistics},
  volume = {47},
  number = {4},
  pages = {2051--2079},
  doi = {10.1214/18-AOS1740},
  langid = {english}
}

@article{fryzlewicz2014wild,
  title = {Wild Binary Segmentation for Multiple Change-Point Detection},
  author = {Fryzlewicz, Piotr},
  year = {2014},
  journal = {The Annals of Statistics},
  volume = {42},
  number = {6},
  eprint = {1411.0858},
  pages = {2243--2281},
  doi = {10.1214/14-AOS1245},
  archiveprefix = {arXiv}
}

@article{horvath2012changepoint,
  title = {Change-Point Detection in Panel Data},
  shorttitle = {Change-Point Detection in Panel Data},
  author = {Horv{\'a}th, Lajos and Hu{\v s}kov{\'a}, Marie},
  year = {2012},
  journal = {Journal of Time Series Analysis},
  volume = {33},
  number = {4},
  pages = {631--648},
  doi = {10.1111/j.1467-9892.2012.00796.x},
  langid = {english}
}

@article{introduction,
  title = {Introduction to {{Data Science}}: {{Data Analysis}} and {{Algorithms}} with {{R}}, {{By Rafael Irrizarry Boca Raton}}, {{FL}}: {{CRC Press}}, 2020. {{Hard}} Cover. Pp. 711.},
  journal = {Biometrics},
  doi = {10.1111/biom.13521}
}

@article{jirak2015uniform,
  title = {Uniform Change Point Tests in High Dimension},
  author = {Jirak, Moritz},
  year = {2015},
  journal = {The Annals of Statistics},
  volume = {43},
  number = {6},
  pages = {2451--2483},
  doi = {10.1214/15-AOS1347},
  langid = {english}
}

@article{kirch2015detection,
  title = {Detection of Changes in Multivariate Time Series with Application to {{EEG}} Data},
  author = {Kirch, Claudia and Muhsal, Birte and Ombao, Hernando},
  year = {2015},
  journal = {Journal of the American Statistical Association},
  volume = {110},
  number = {511},
  pages = {1197--1216},
  doi = {10.1080/01621459.2014.957545},
  langid = {english}
}

@article{kolda2009tensor,
  title = {Tensor Decompositions and Applications},
  author = {Kolda, Tamara G. and Bader, Brett W.},
  year = {2009},
  journal = {SIAM Review},
  volume = {51},
  number = {3},
  pages = {455--500},
  doi = {10.1137/07070111X},
  langid = {english}
}

@article{matrix,
  title = {Matrix {{Completion}}, {{Counterfactuals}}, and {{Factor Analysis}} of {{Missing Data}}},
  journal = {Journal of the American Statistical Association},
  doi = {10.1080/01621459.2021.1967163}
}

@article{ombao2005slex,
  title = {{{SLEX}} Analysis of Multivariate Nonstationary Time Series},
  author = {Ombao, Hernando and {von Sachs}, Rainer and Guo, Wensheng},
  year = {2005},
  journal = {Journal of the American Statistical Association},
  volume = {100},
  number = {470},
  pages = {519--531},
  doi = {10.1198/016214504000001448},
  langid = {english}
}

@article{preuss2015detection,
  title = {Detection of Multiple Structural Breaks in Multivariate Time Series},
  author = {Preuss, Philip and Puchstein, Ruprecht and Dette, Holger},
  year = {2015},
  journal = {Journal of the American Statistical Association},
  volume = {110},
  number = {510},
  pages = {654--668},
  doi = {10.1080/01621459.2014.920613},
  langid = {english}
}

@article{schroder2019fresped,
  title = {{{FreSpeD}}: {{Frequency-specific}} Change-Point Detection in Epileptic Seizure Multi-Channel {{EEG}} Data},
  shorttitle = {Fresped},
  author = {Schr{\"o}der, Anna Louise and Ombao, Hernando},
  year = {2019},
  journal = {Journal of the American Statistical Association},
  volume = {114},
  number = {525},
  pages = {115--128},
  doi = {10.1080/01621459.2018.1476238},
  langid = {english}
}

@article{small,
  title = {Small Deviation Estimates for the Largest Eigenvalue of {{Wigner}} Matrices. ({{arXiv}}:2112.12093v1 [Math.{{PR}}])},
  journal = {arXiv Statistics Theory},
  doi = {arXiv:2112.12093v1}
}

@article{sparsity,
  title = {Sparsity {{Concepts}} and {{Estimation Procedures}} for {{High Dimensional Vector Autoregressive}}  {{Models}}},
  journal = {Journal of Time Series Analysis},
  doi = {10.1111/jtsa.12586}
}

@article{stevenson2019dataset,
  title = {A Dataset of Neonatal {{EEG}} Recordings with Seizure Annotations},
  author = {Stevenson, N. J. and Tapani, K. and Lauronen, L. and Vanhatalo, S.},
  year = {2019},
  journal = {Scientific Data},
  volume = {6},
  number = {1},
  pages = {1--8},
  publisher = {Nature Publishing Group},
  doi = {10.1038/sdata.2019.39},
  copyright = {2019 The Author(s)},
  langid = {english}
}

@article{sun2017provable,
  title = {Provable Sparse Tensor Decomposition},
  author = {Sun, Will Wei and Lu, Junwei and Liu, Han and Cheng, Guang},
  year = {2017},
  journal = {Journal of the Royal Statistical Society: Series B (Statistical Methodology)},
  volume = {79},
  number = {3},
  pages = {899--916},
  doi = {10.1111/rssb.12190},
  langid = {english}
}

@article{wang2018high,
  title = {High Dimensional Change Point Estimation via Sparse Projection},
  author = {Wang, Tengyao and Samworth, Richard J.},
  year = {2018},
  journal = {Journal of the Royal Statistical Society: Series B (Statistical Methodology)},
  volume = {80},
  number = {1},
  pages = {57--83},
  doi = {10.1111/rssb.12243},
  langid = {english}
}

@article{wang2021optimal,
  title = {Optimal Covariance Change Point Localization in High Dimensions},
  author = {Wang, Daren and Yu, Yi and Rinaldo, Alessandro},
  year = {2021},
  journal = {Bernoulli},
  volume = {27},
  number = {1},
  pages = {554--575},
  doi = {10.3150/20-BEJ1249}
}

@article{wu2005nonlinear,
  title = {Nonlinear System Theory: {{Another}} Look at Dependence},
  shorttitle = {Nonlinear System Theory},
  author = {Wu, Wei Biao},
  year = {2005},
  journal = {Proceedings of the National Academy of Sciences},
  volume = {102},
  number = {40},
  pages = {14150--14154},
  doi = {10.1073/pnas.0506715102},
  langid = {english}
}

@article{yuan2013truncated,
  title = {Truncated Power Method for Sparse Eigenvalue Problems},
  author = {Yuan, Xiao-Tong and Zhang, Tong},
  year = {2013},
  journal = {Journal of Machine Learning Research},
  volume = {14},
  number = {Apr},
  pages = {899--925}
}

@article{zhang2021convergence,
  title = {Convergence of Covariance and Spectral Density Estimates for High-Dimensional Locally Stationary Processes},
  author = {Zhang, Danna and Wu, Wei Biao},
  year = {2021},
  journal = {The Annals of Statistics},
  volume = {49},
  number = {1},
  pages = {233--254},
  doi = {10.1214/20-AOS1954}
}

\end{document}